\documentclass[aps,prd,twocolumn,showpacs,floatfix,preprintnumbers,amsfont,amsmath,amssymb,nofootinbib,superscriptaddress]{revtex4-1}

\usepackage{color}
\usepackage{bbold}
\usepackage[colorlinks=true,linktocpage=true,linkcolor=colorLink,citecolor=colorCite,urlcolor=colorURL]{hyperref}
\usepackage[normalem]{ulem}
\usepackage{lipsum}
\usepackage{url}
\usepackage{float}
\usepackage[table]{xcolor}
\usepackage{graphicx}
\usepackage[font=small, justification=raggedright, format=plain]{caption}
\usepackage{subcaption}
\usepackage{mathtools}
\usepackage{bm}
\usepackage{siunitx}
\usepackage{multirow}
\definecolor{colorLink}{rgb}{0.9,0,0} % red
\definecolor{colorCite}{rgb}{0,0.7,0} % green
\definecolor{colorURL} {rgb}{0,0,0.8} % navy

\usepackage{array,etoolbox}
\preto\tabular{\setcounter{magicrownumbers}{-1}}
\newcounter{magicrownumbers}

\colorlet{RED}{red}

\def\pastro{p_{\rm astro}}
\def\msun{{\rm M_{\odot}}}
\newcommand{\mcal}[1]{\mathcal{#1}}

\newcommand{\sk}[1]{}

\newcommand{\be}{\begin{equation}}
\newcommand{\ee}{\end{equation}}
\newcommand{\ba}{\begin{eqnarray}}
\newcommand{\ea}{\end{eqnarray}}

\newcommand{\apjl}{Astrophys. J. Lett.}

\newcommand{\mnras}{Mon. Not. R. Astron. Soc.}

\newcommand{\neviasnewtot}{6}
\newcommand{\neviasnew}{6}
\newcommand{\neviasconf}{17}

\newcommand{\neviastot}{23}
\newcommand{\nevfalseposias}{2} % number of fasle positives using Sum_i (1 - pastro_i) over the new events
\newcommand{\nevlvc}{35}

\newcommand{\nevlvchl}{30}

\newcommand{\naddogc}{7}

\allowdisplaybreaks
\begin{document}

\title{New binary black hole mergers in the LIGO--Virgo O3b data}

\author{Ajit Kumar Mehta}
\email{ajit\_mehta@ucsb.edu}
\affiliation{\mbox{Department of Physics, University of California at Santa Barbara, Santa Barbara, CA 93106, USA}}
\author{Seth Olsen}
\affiliation{\mbox{Department of Physics, Princeton University, Princeton, NJ 08540, USA}}
\author{Digvijay Wadekar}
\affiliation{\mbox{School of Natural Sciences, Institute for Advanced Study, 1 Einstein Drive, Princeton, NJ 08540, USA}}
\author{Javier Roulet}
%\email{jroulet@ucsb.edu}
\affiliation{\mbox{TAPIR, Walter Burke Institute for Theoretical Physics, California Institute of Technology, Pasadena, CA 91125, USA}}
\author{Tejaswi Venumadhav}
\affiliation{\mbox{Department of Physics, University of California at Santa Barbara, Santa Barbara, CA 93106, USA}}
\affiliation{\mbox{International Centre for Theoretical Sciences, Tata Institute of Fundamental Research, Bangalore 560089, India}}
\author{Jonathan Mushkin}
\affiliation{\mbox{Department of Particle Physics \& Astrophysics, Weizmann Institute of Science, Rehovot 76100, Israel}}
\author{Barak Zackay}
\affiliation{\mbox{Department of Particle Physics \& Astrophysics, Weizmann Institute of Science, Rehovot 76100, Israel}}
\author{Matias Zaldarriaga}
\affiliation{\mbox{School of Natural Sciences, Institute for Advanced Study, 1 Einstein Drive, Princeton, NJ 08540, USA}}

\date{\today}

%%%%%%%%%%%%%%%%%%%%%%%%%%%%%%%%%%%%%%%%%%%%%%%%%%%%%%%%%%%%%%%%%%%%%%%%%%%%%%%

\begin{abstract}

We report the detection of \neviasnewtot{} new candidate binary black hole (BBH) merger signals in the publicly released data from the second half of the third observing run (O3b) of advanced LIGO and advanced Virgo. The LIGO--Virgo--KAGRA (LVK) collaboration reported \nevlvc{} compact binary coalescences (CBCs) in their analysis of the O3b data \cite{lvc_gwtc3_o3_ab_catalog_2021}, with \nevlvchl{} BBH mergers having coincidence in the Hanford and Livingston detectors. We confirm \neviasconf{} of these for a total of \neviastot{} detections in our analysis of the Hanford--Livingston coincident O3b data. We identify candidates using a search pipeline employing aligned-spin quadrupole-only waveforms. Our pipeline is similar to the one used in our O3a coincident analysis \cite{ias_o3a_pipeline2022}, except for a few improvements in the veto procedure and the ranking statistic, and we continue to use an astrophysical probability of one half as our detection threshold, following the approach of the LVK catalogs. Most of the new candidates reported in this work are placed in the upper/lower-mass gap of the black hole (BH) mass distribution.  We also identify a possible neutron star-black hole (NSBH) merger. We expect these events to help inform the black hole mass and spin distributions inferred in a full population analysis.

\end{abstract}

\maketitle

\section{Introduction}

The Advanced LIGO \cite{advancedLIGO_aasi2015} and Advanced Virgo \cite{advancedVIRGO_2014yos} detectors have now been joined by KAGRA \cite{KAGRA:2020tym} to form the LIGO--Virgo--KAGRA (LVK) collaboration, which recently released data from the second half of the third observing run (O3b) to the public \cite{GWOSC}. The LVK detected gravitational waves (GW) from ~\nevlvc~ compact binary coalescence (CBC) events including two probable neutron star--black hole (NSBH) mergers \cite{lvc_gwtc3_o3_ab_catalog_2021}. 32 of these were found to be coincident in the Hanford--Livingston (HL) detectors, among which 30 were reported to be binary black hole (BBH) signals. \citet{nitz_4ogc_o3_ab_catalog_2021} independently produced the 4-OGC catalog, which included an additional ~\naddogc~ BBH mergers that were not in the LVK catalog. {However, it also missed 8 events reported in GWTC-3 that have a probability of astrophysical origin ($\pastro$) greater than 0.5. This emphasizes that different pipelines identify events near the threshold with different significance.}

\begin{figure*}
    \centering
    \includegraphics[height=.22\textheight]{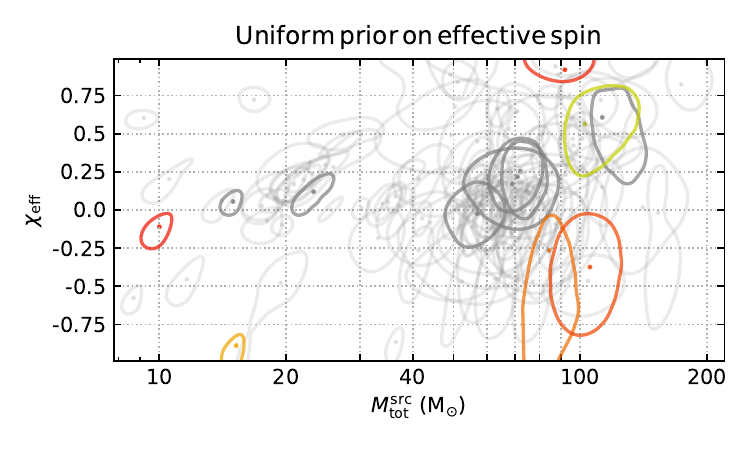}\hfill
    \includegraphics[height=.22\textheight]{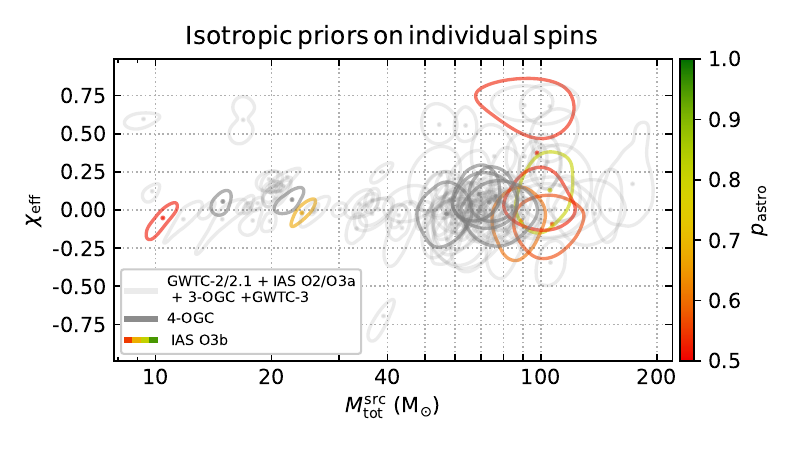}
    \vspace{-0.3cm}
    \caption{The source-frame total mass and effective spin for the BBH events detected so far from all the search pipelines across the first three observing runs of LVK. Posterior contours enclose $50\%$ of the probability and the median is represented by a dot. For parameter inference, we use the \texttt{IMRPhenomXPHM} waveform approximant \cite{xphm_pratten2020} and a prior that is uniform in detector-frame constituent masses and comoving volume-time (redshifts based on Planck15 results \cite{cosmology_planck2015}). Contours for the \neviasnew{} new events are colored by their $\pastro$ values, and the events appearing in previous IAS, OGC and LVK catalogs up to 3-OGC and GWTC-3 \cite{O1catalog_LVC2016, ias_pipeline_o1_catalog_new_search_prd2019, gwtc1_o2catalog_LVC2018, ias_o2_pipeline_new_events_prd2020, lvc_o3a_gwtc2_catalog_2021, lvc_o3a_deep_gwtc2_1_update_2021, ias_o3a_pipeline2022, lvc_gwtc3_o3_ab_catalog_2021} are transparent gray. The additional (7) events reported in 4-OGC catalog  \cite{nitz_4ogc_o3_ab_catalog_2021} from O3b are shown with solid gray color. Though our search only covers events with Hanford--Livingston coincidence triggers, we include all BBH events declared in the LVK and OGC catalogs through O3b in this figure. \emph{Left panel:} Results obtained with uniform prior on the effective spin (default prior used in this work). \emph{Right panel:} Results with isotropic priors on individual spins, as used in the GWTC-3 and 4-OGC catalogs.}
    %\jay{isotropic spelling in fig title}
    %\barak{I suggest putting titles to the plots, to make them standalone copy pastable, as well as easier to grasp when you skim the paper}
    \label{fig:population}
\end{figure*}

{In this work, we perform a search analysis on the O3b HL coincident data (excluding Virgo)} using a pipeline closely resembling the one used in \citet{ias_o3a_pipeline2022} (see \citet{ias_pipeline_o1_catalog_new_search_prd2019} for an overview of the pipeline). {We report detections of a total of \neviastot{} compact binary mergers, \neviasnewtot{} of which are reported for the first time here. The search did not recover 13 events reported in GWTC-3 and the additional 7 events reported in 4-OGC (more details follow later in the introduction, with a detailed description provided in section \ref{sec:previous_events})}.
We show these detections in Fig.~\ref{fig:population} alongside the 90 CBC signals detected by LVK, additional events detected in 4-OGC, and the events previously detected by our pipeline in O1-O3a with $\pastro$ greater than one half \cite{O1catalog_LVC2016, gwtc1_o2catalog_LVC2018, NitzCatalog_1-OGC_o1_2018, NitzCatalog_2-OGC_o2_2020,lvc_o3a_gwtc2_catalog_2021, lvc_o3a_deep_gwtc2_1_update_2021, ias_pipeline_o1_catalog_new_search_prd2019, ias_o2_pipeline_new_events_prd2020, ias_o3a_pipeline2022}.

Understanding the mass and spin distributions of the astrophysical BBH population is one of the major goals of GW astronomy. As the number of events in the catalog increases, statistically inferred measurements of the population parameters become more and more precise \cite{ias_o3a_population_analysis_prd2021roulet, lvc_o3a_population_properties_2021}. These improved measurements help us to constrain BBH formation channel rates, and they allow us to probe deeper into the fundamental physics of black holes (BHs) and even cosmology \cite{coherent_mode_stacking_BH_spectroscopy_Yang2017pretorius, constraints_on_beyondGR_Perkins2021yunes, measure_hubble_from_population_Abbott2021lvc, unmodeled_mode_stacking_measure_HM_from_population_OBrien2019costa, testing_large_scale_GR_important_for_pop_mass_dist_Ezquiaga2021}.

One of the standard BBH formation channels consists of binary stars that evolve in isolation. Stellar evolution models predict that, if massive enough, stars can go through pulsational pair-instability (PPISN) and pair-instability supernovae (PISN). In particular, the latter can completely destroy the stars, leaving no remnant at the end of the evolution. The standard prediction is that this will lead to an ``upper mass gap" (UMG) in which the astrophysical distribution drops sharply for BH masses between $\sim 45\,\msun$ and $\sim 135\,\msun$\cite{umg_pre_pairinstability_early_fowler1964ApJS, umg_Barkat1967PairInstability, umg_another_old_pairinstability_bond1984ApJ, umg_HegerWoosley2002massgap, umg_Woosley2007pairinstabilitySN, umg_Woosley2017ppisn, umg_Farmer2019LowerEdgeBHMassGap, umg_in_collapse_simulations_chen2014, umg_yoshida2016_PPI_simulations, umg_pair_instability_mass_loss_Belczynski2016}. \sk{There also exists an empirical ``lower mass gap'' (LMG) between $\sim 2\,\msun$ and $\sim 5\,\msun$ corresponding to the region between the maximum NS mass \cite{max_NS_mass_est_from_EOS_space_alsing2018berti} and the lowest observed BH mass from stellar collapse} Local observations of BHs also hint at the existence of an empirical ``lower mass gap'' (LMG) between $\sim 2\,\msun$ and $\sim 5\,\msun$ corresponding to the region between the maximum NS mass \cite{max_NS_mass_est_from_EOS_space_alsing2018berti} and the lowest observed BH mass \cite{lmg_OBS_EM_stellarBH_Mdist_from_few_xrayNSBH_Bailyn1998, lmg_OBS_EM_BHdistInfer_from_16xrayNSBH_ozel2010, lmg_OBS_EM_BHdist_from_15xrayNSBHrocheOver_Farr2011ilya, lmg_model_3methods_1gap_fastSNtime_2012, lmg_bridging_need_NSBHfeature_Farah2021fishbach_holz, lmg_from_CCSNe_simulations_Liu2020-21sun} (note that according to \citet{max_NS_mass_est_from_EOS_space_alsing2018berti}, the maximum mass of the NS can vary between $[2, 2.6]\,\msun$). Detecting BHs in these gaps can have several implications, e.g., detecting a BH in the UMG suggests an alternative to the isolated stellar evolution mechanism for its formation, such as hierarchical mergers in dense stellar environments (the ``dynamical formation" channel) \cite{star_cluster_dynamical_v_isolated_rates_mapelli2020, metallicity_in_young_star_clusters2020a, young_star_clusters_populating_mass_gap2020a, cluster_hierarchical_metallicity_spin_mapelli2021, young_star_clusters_heavy_remnants2021, agn_accretion_disk_merger_population2020a, agn_bbh_population_chieff_q_simulation_mckernan_ford2019, mass_gap_agn_bbh_mergers2021, merger_rates_agn_Tagawa_2020a, hierarchical_mergers_agn_kocsis2019, ishibashi_rate_estimate_merger_agn2020a}) and/or they can make us revisit the problem of stellar collapse \cite{Farmer:2019jed, Marchant:2020haw, Mehta:2021fgz}.  
As we will see later, among the new events presented here we do find sources whose constituents lie in these gaps (see Table~\ref{tab:signalsFoundParsPESourceFrame}). 
Including such events in the population analysis could \sk{significantly} impact the inferred BH mass distribution \cite{lvc_o3a_population_properties_2021, ias_o3a_population_analysis_prd2021roulet}. 

In addition to the masses, the BH spins can also be measured from GW signals. The best measured parameter, however, is a particular combination of the masses and spins called the effective spin:
\begin{align} \label{eq:chieff_def}
\chi_{\rm eff} = \frac{m_1 \chi_{1,z} + m_2 \chi_{2,z}}{m_1 + m_2}\,.
\end{align}
where $m_i$ are the masses and $\chi_{i,z}$ are the dimensionless spins projected on the orbital angular momentum. Together with the measurement of the masses, measuring $\chi_{\rm eff}$ could give us more information about the underlying formation channels. For example, as the total mass of the binaries increases, highly spinning systems with $\chi_{\rm{eff}} < 0$ (anti-aligned) become increasingly difficult to form in the isolated binary evolution channel even under a broad range of BH natal kicks \cite{need_dynamical_to_misalign_orbit_and_spin_Rodriguez2016}. Similar to our previous work \cite{ias_o3a_pipeline2022}, we employ a spin prior which is uniform in $\chi_{\rm{eff}}$. This prior is also useful if a reader wants to reweight these results to another prior of their choice, e.g., a prior that is isotropic in the individual spins such as the one used in the GWTC-3 \cite{lvc_gwtc3_o3_ab_catalog_2021} and 4-OGC \cite{nitz_4ogc_o3_ab_catalog_2021} catalogs . Though we use uniform in $\chi_{\rm{eff}}$ as our default spin prior for our parameter estimation runs, we perform an additional set of parameter estimation runs to compare against the isotropic spin prior results\sk{ wherever required}.

% As mentioned in \citet{ias_o3a_pipeline2022}, the individual parameter estimation (PE) results as well as the population results could be altered based on what prior we choose. Prior assumptions even go into the estimation of the significance of the events which determines whether an event should be included in the population analysis or not. On the other hand, nobody knows what the true prior is, and thus it is not possible to completely escape from this. One could use astrophysically motivated priors like isotropic in spin orientations motivated by the dynamical formation channel~\footnote{This prior is used in other catalogs [CITE]}. However, it is quite possible that we witness a signal in the parameter space where this prior leads to inadequate samples which could then lead to the failure of the reweighting procedure afterward. Keeping this in mind, following \cite{ias_o3a_pipeline2022} we here also use a uniform prior on the best measured combination of mass and spin variables (i.e., the effective spin) \cite{flat_chieff_prior_o1event1216_formation_channels2019}. In what follows we will see that some of our results do depend on which of these two priors is used in the PE analysis.    

In this work, we follow-up on our previous study (\citet{ias_o3a_pipeline2022}) which looked for new events in the Hanford--Livingston (HL) data from the first half of the third observing run (O3a), and analyze the data from the second half of the third observing run (O3b) to complete an O3 catalog with a similar pipeline (see Section \ref{sec:changes} for a description of the changes). We detect \neviasnew{} new BBH candidate events under the criterion that the astrophysical probability $\pastro$ is at least one half (which is also the threshold used in the GWTC-3 and 4-OGC catalogs). By simply summing the complements of the $\pastro$ values assigned to these events, we expect that roughly \nevfalseposias{} of the \neviasnew{} events may be noise transients rather than astrophysical signals (note, however, that the estimates of $\pastro$ depend on multiple factors including the astrophysical prior used, and that the dependence is strongest in the case of near-threshold detections such as these).

Our pipeline confirmed the detections of \neviasconf{} out of the 30 HL coincident BBH mergers reported in the GWTC-3 catalog. 
When we investigated our pipeline's performance on the remaining events in the LVK catalog, we found that 6 of them were marginal in our catalog, i.e., recovered with a $\pastro$ lower than one half, while 6 were vetoed for failing one or more of our signal consistency and excess power tests, and one event was missed. 
Most of the marginally recovered events were detected by only one of the LVK search pipelines, while three of the vetoed events show signs of physical effects like spin-precession and higher harmonics that were not modeled in our template bank. We also note that four of the vetoed events were flagged as having data-quality issues in the GWTC-3 catalog paper. Finally, we also checked that 4 of the \neviasnew{} new candidates reported in this work were also present in the list of GWTC-3 sub-threshold candidates \cite{GWTC-3Marginal}. 
%\jay{maybe add a comparison table in the appendix?}. 

The new GW sources reported here could come from a range of astrophysical scenarios. Three events have their primary (i.e. more massive) constituents confidently placed in the upper mass gap.  One event has its secondary (i.e. less massive) constituent confidently placed in the lower mass gap, i.e. $ 2\,\msun \lesssim m_{2}^{\rm{src}} \lesssim 5\,\msun$. If we allow for the possibility that the maximum mass of the NS can reach up to $\sim 2.6\,\msun$ \cite{max_NS_mass_est_from_EOS_space_alsing2018berti},  this event could also be a neutron star-black hole (NSBH) signal at $90\%$ confidence (note however that we did not perform a dedicated NSBH search). There is also a particular low-mass event which, under uniform prior on effective spin, seems to have strongly anti-aligned spins ($\chi_{\rm{eff}} \leq -0.8$ at $90\%$ confidence; see the discussion in the text for caveats). 

We find that for at least two of the new events, detailed parameter estimation shows some evidence for effects such as higher-order modes (HM) and spin-precession, and also some information from the VIRGO data, which suggests that their significance could be potentially improved by including these elements in future improved searches (our current search of the Hanford--Livingston coincident data uses aligned-spin templates that only have the fundamental ($\ell = |m| = 2$) harmonic). 

% \jay{Here or somewhere else, you can mention that in our new HM search \cite{Wad23_Events}, we indeed recover GW200109\_195634 with a higher SNR \{$\rho^2_{22}, \rho^3_{33}, \rho^2_{44} = 63.7, 5.0,8.7$\}}

As noted earlier, some loud events in the LVK catalog that were vetoed by our pipeline also showed evidence for these physical effects, which additionally motivates the development of improved pipelines incorporating them. {We note that the LVK catalog was also produced using aligned-spin templates limited to the $\ell = |m| = 2$ harmonic, as in our case. However, our veto tests are comparatively more stringent, making it crucial for us to search for signals showing hints of HM or precession using templates that explicitly include these effects. }

The rest of the paper is organized as follows: in \S \ref{sec:changes} we review changes to the IAS pipeline between the O3a and O3b analyses. In \S \ref{sec:events} we discuss the BBH mergers first reported in this work (see Table~\ref{tab:signalsFoundParsPESourceFrame}). In \S \ref{sec:previous_events} we report our results for events already included in GWTC-3 \cite{lvc_gwtc3_o3_ab_catalog_2021} and 4-OGC \cite{nitz_4ogc_o3_ab_catalog_2021}, noting differences (see Table~\ref{tab:lvc events}). We summarize the results in \S \ref{sec:conclusions} and discuss the astrophysical implications of the new events. In Appendix \ref{Appendix:pastro}, we show the computation of the astrophysical probability, $\pastro$ and also discuss certain details of our veto procedure.

Corner plots of posterior distributions for new events can be found in Appendix \ref{Appendix:posteriors}, with PE samples publicly available at \url{https://github.com/AjitMehta/new_BBH_mergers_O3b_IAS_pipeline}.

%%%%%%%%%%%%%%%%%%%%%%%%%%%%%%%%%%%%%%%%%%%%%%%%%%%%%%%%%
%%%%%%%%%%%%%%%%%%%%%%%%%%%%%%%%%%%%%%%%%%%%%%%%%%%%%%%%%
\section{Changes to the O3a analysis pipeline}
\label{sec:changes}
Our analysis pipeline is very similar to the version used in our recent analysis of the Hanford--Livingston coincident O3a data \cite{ias_o3a_pipeline2022}, but we make changes to two components of the pipeline detailed below in this section. Once we have the ranked lists of coincident foreground and background candidates, we calculate the candidates' $\pastro$ values using the same procedure as in Appendix~\ref{Appendix:pastro} of \citet{ias_o3a_pipeline2022}. Note that, unlike the results reported in the previous IAS papers, we also report, for each event, an overall inverse false-alarm rate (IFAR) combined across our template banks (assuming an equal astrophysical rate density for each bank) in addition to the IFAR computed within the respective bank. 

\subsection{Signal-consistency veto tests: } In the matched-filtering step of our pipeline, we collect and save triggers above a particular threshold on the signal-to-noise ratio (SNR; we use $\rho^2 \gtrsim$ 15 for the template banks for the most massive BBHs, where $\rho$ is the single-detector SNR; the exact threshold varies with the bank and is typically higher for banks for less massive BBHs). In our previous studies, we only performed our signal-consistency veto tests on the subset of these triggers that passed a higher threshold on their SNR ($\rho^2\geq 30$). Ideally, one should perform the veto tests on all the candidates irrespective of their SNR, but we need to pick a higher threshold as the veto tests are computationally expensive (they take roughly a minute per trigger per CPU in a single detector). This choice can lead to some spurious triggers (i.e., ones that should ideally be vetoed) that are below the SNR threshold entering our final ranked lists (this happens to both the coincident foreground and background triggers). 

To ameliorate this, we use the following procedure. We first construct our ranked lists using the same $\rho^2$ veto threshold as before.
    We then collect high-ranking candidates with SNR below the veto threshold (which had been omitted from our veto tests before), and screen them through the veto tests (we do this procedure on $\sim 10000$ top candidates from our background and coincident lists). We make our final ranked candidate lists by removing any candidates which failed at least one veto test. There is another slight change we make to the order in which the veto procedure is applied to the triggers in the pipeline compared to the O3a pipeline, which we discuss in Appendix \ref{Appendix:pastro}.  
    
    %\sout{These modifications, though minor, are roughly observed to mildly improve the sensitivity of our pipeline.}

    % procedure in two steps. The first step involves doing the same as previously, i.e., we select the triggers based on the usual thresholds and carry out the veto tests. The second step involves picking up top $\sim 10000$ triggers from the background again and applying the veto tests on the triggers which skipped due to the threshold criteria \footnote{Similar procedure is followed for the foreground.}. Our motivation to do this came from the observation that some triggers below our usual SNR threshold too get vetoed, improving the background.  We also change the order in which the veto procedure is applied to the triggers along the pipeline (see Appendix \ref{Appendix:pastro} for detailed discussion).

\subsection{Ranking statistic} 
Our statistic to rank the triggers performs two functions: $(i)$ it up-weights triggers which have good coherence among multiple detectors, ($ii$) it down-weights triggers in individual detectors based on the probability with which similar triggers occur in the background % distribution (we obtain the background using the method of timeslides). 
(for a detailed discussion on the ranking statistic used in the pipeline, see Appendix \ref{Appendix:pastro} and Ref.~\cite{Wad23_Pipeline}). Corresponding to the second function, our ranking statistic includes a term: $-2\log P(\rho^2 | N)$ -- this is the probability of obtaining a trigger with squared SNR $\rho^2$ from the background for a given detector (in our previous papers, we labelled this term as the $\textit{rank function}$ (denoted by $\tilde{\rho}$), see section~J of Ref.~\cite{ias_pipeline_o1_catalog_new_search_prd2019}). We measure $P(\rho^2 | N)$ by constructing a histogram of the SNR values of background triggers (obtained using timeslides) and correcting for the incompleteness introduced by our cuts. 

\bgroup
\def\arraystretch{1.5}
\begin{table*}
    \centering
    \caption{New events with $p_{\rm astro} > 0.5$. The parameter values correspond to the median and $90\%$ uncertainties and $\ln \mathcal{L}_{\rm{max}}$ denotes the maximum log likelihood from parameter estimation (PE) runs. While our search templates only include aligned-spin (2,2) mode (based on the \texttt{IMRPhenomD} approximant), PE was done with \texttt{IMRPhenomXPHM} waveform which additionally includes effects of precession and higher-harmonics: (2,1), (3,3) and (4,4) modes. The PE also uses Virgo data when available, unlike the search. $\rho_{\rm{H,L}}$ denote the SNR of the triggers in the Hanford and Livingston detectors. The IFAR column shows the IFAR within the bank and the one combined across the chirp-mass banks assuming an equal astrophysical rate density for each chirp-mass bank.}
    \begin{tabular}{|c|c|ccccc|cc|cc|c|}
        \hline
        \multirow{2}{*}{Event} & \multirow{2}{*}{Bank} &\multicolumn{5}{c|}{PE} & \multirow{2}{*}{$\rho_{\rm H}$}  & \multirow{2}{*}{$\rho_{\rm L}$} &\multicolumn{2}{c|}{IFAR (yr)} & \multirow{2}{*}{$p_\mathrm{astro}$} \\
        \cline{3-7} \cline{10-11}
         & & $m_1^{\textrm{src}} {\scriptstyle( \rm M_\odot)}$ & $m_2^{\textrm{src}} {\scriptstyle(\rm M_\odot)}$ & $\chi_{\rm eff}$ & $z$ & $\ln \mathcal{L}_\mathrm{max}$ & & &  within bank $|$ & $|$ overall  &
         \\ 
         \hline 

\rowcolor{gray!30} GW200109\_195634$\footnote{This event's significance could be potentially improved in future searches in which Virgo data is also included in the coincident analysis.}$ & \texttt{(5, 4)} & $69_{-19}^{+24}$ & $48_{-17}^{+22}$ & $0.5_{-0.8}^{+0.3}$ & $1.0_{-0.4}^{+0.7}$ & $42.0$ & $5.5$ & $5.8$ & $4.4$ & $0.93$ & $0.81$ \\
GW191228\_085854$\footnote{This event has bi-modal posterior distributions in the intrinsic parameters. The values quoted here correspond to the first mode. See \ref{event_GW191228} for further details.}$  & \texttt{(1, 0)} & $9.7_{-1.5}^{+4.2}$ & $5.5_{-1.4}^{+0.9}$ & $-0.89_{-0.10}^{+0.42}$ & $0.14_{-0.04}^{+0.05}$ & $43.2$ & $5.2$ & $7.3$ & $1.8$ & $0.25$ & $0.67$ \\
\rowcolor{gray!30} GW200225\_075134 & \texttt{(3, 1)} & $51_{-11}^{+17}$ & $37_{-11}^{+13}$ & $-0.5_{-0.4}^{+0.9}$ & $0.6_{-0.3}^{+0.4}$ & $43.0$ & $4.5$ & $7.0$ & $0.95$ & $0.15$ & $0.60$ \\
GW191117\_023843 & \texttt{(5, 4)} & $62_{-15}^{+26}$ & $44_{-15}^{+17}$ & $-0.4_{-0.4}^{+0.6}$ & $0.9_{-0.4}^{+0.5}$ & $33.3$ & $6.5$ & $4.7$ & $0.52$ & $0.12$ & $0.56$ \\
\rowcolor{gray!30} GW200210\_100022 & \texttt{(5, 2)} & $74_{-17}^{+15}$ & $17_{-4}^{+29}$ & $0.92_{-0.42}^{+0.07}$ & $1.3_{-0.4}^{+0.7}$ & $34.2$ & $4.9$ & $6.2$ & $0.45$ & $0.097$ & $0.52$ \\
GW200316\_235947 & \texttt{(0, 0)} & $6.4_{-1.4}^{+4.9}$ & $3.6_{-1.3}^{+1.0}$ & $-0.11_{-0.13}^{+0.34}$ & $0.17_{-0.05}^{+0.05}$ & $38.9$ & $6.1$ & $5.7$ & $0.59$ & $0.092$ & $0.52$ \\

% \rowcolor{gray!30} GW200109\_195634$\footnote{This event's significance could be potentially improved in future searches in which Virgo data is also included in the coincident analysis.}$ & \texttt{(5, 4)} & $69_{-19}^{+24}$ & $48_{-17}^{+22}$ & $0.5_{-0.8}^{+0.3}$ & $1.0_{-0.4}^{+0.7}$ & $42.0$ & $30.6$ & $33.6$ & $4.4$ & $0.93$ & $0.81$ \\
% GW191228\_085854~\footnote{This event has bi-modal posterior distributions in the intrinsic parameters. The values quoted here correspond to the first mode. See \ref{event_GW191228} for further details.}  & \texttt{(1, 0)} & $9.7_{-1.5}^{+4.2}$ & $5.5_{-1.4}^{+0.9}$ & $-0.89_{-0.10}^{+0.42}$ & $0.14_{-0.04}^{+0.05}$ & $43.2$ & $26.9$ & $54.0$ & $1.8$ & $0.25$ & $0.67$ \\
% \rowcolor{gray!30} GW200225\_075134 & \texttt{(3, 1)} & $51_{-11}^{+17}$ & $37_{-11}^{+13}$ & $-0.5_{-0.4}^{+0.9}$ & $0.6_{-0.3}^{+0.4}$ & $43.0$ & $20.0$ & $49.4$ & $0.95$ & $0.15$ & $0.60$ \\
% GW191117\_023843 & \texttt{(5, 4)} & $62_{-15}^{+26}$ & $44_{-15}^{+17}$ & $-0.4_{-0.4}^{+0.6}$ & $0.9_{-0.4}^{+0.5}$ & $33.3$ & $41.8$ & $21.9$ & $0.52$ & $0.12$ & $0.56$ \\
% \rowcolor{gray!30} GW200210\_100022 & \texttt{(5, 2)} & $74_{-17}^{+15}$ & $17_{-4}^{+29}$ & $0.92_{-0.42}^{+0.07}$ & $1.3_{-0.4}^{+0.7}$ & $34.2$ & $23.6$ & $37.9$ & $0.45$ & $0.097$ & $0.52$ \\
% GW200316\_235947 & \texttt{(0, 0)} & $6.4_{-1.4}^{+4.9}$ & $3.6_{-1.3}^{+1.0}$ & $-0.11_{-0.13}^{+0.34}$ & $0.17_{-0.05}^{+0.05}$ & $38.9$ & $36.7$ & $32.7$ & $0.59$ & $0.092$ & $0.52$ \\

         \hline
    \end{tabular}
    \label{tab:signalsFoundParsPESourceFrame}
\end{table*}
\egroup

In the previous iterations of our search pipeline, we combined background triggers from all templates in a given subbank to construct this histogram and estimate the rank function. In reality, $\log P(\rho^2 | N)$ can have different values even for different templates in a given subbank, i.e., the background can be strongly template-dependent, see e.g., \cite{LIGO_O1, CoherentScore, Wad23_Pipeline}.
In order to optimize our analysis, we follow the technique in Ref.~\cite{Wad23_Pipeline}, where we first cluster the templates within each sub-bank based on their sensitivity to glitches (as quantified empirically by the relative fraction of loud and faint triggers belonging to each template). We then separately construct the histograms of background triggers for each template group, and estimate a rank function for triggers belonging to templates within that group. A detailed discussion on this point will be provided in our subsequent paper \cite{Wad23_Pipeline}.\\

We expect the senstivity of our pipeline to increase due to these two improvements. We plan to quantify this increase in terms of the volume-time ($VT$) based on the performance of our pipeline on recovery of injected signals in the data in a future paper. 

% \jay{I will rewrite this sub-section (Jay).}
% Our statistic to rank the triggers includes the term: $-2\log P(\rho^2 | N)$, i.e. the probability of obtaining a trigger with squared SNR 
% %\seth{of at least}
% $\rho^2$ from the background for a given detector (note that this term has been labelled as the $\textit{rank function}$ in our previous studies, see section~J of \cite{ias_pipeline_o1_catalog_new_search_prd2019}). Note that the background trigger distribution, and thus the rank function, is a function of templates. Previously, we constructed a single rank function for each sub-bank, which was \sk{indeed an}\seth{a rough} approximation. To make it more precise, in this work we make $n$ groups of templates in each sub-bank depending on the similarity of templates towards what we call ``glitchiness", and then construct a rank function for each of these groups. A detailed discussion on this will be provided in our subsequent paper \cite{Wad23_Pipeline}.\\

%%%%%%%%%%%%%%%%%%%%%%%%%%%%%%%%%%%%%%%%%%%%%%%%%%%%%%%%%
%%%%%%%%%%%%%%%%%%%%%%%%%%%%%%%%%%%%%%%%%%%%%%%%%%%%%%%%%
%%%%%%%%%%%%%%%%%%%%%%%%%%%%%%%%%%%%%%%%%%%%%%%%%%%%%%%%%
%%%%%%%%%%%%%%%%%%%%%%%%%%%%%%%%%%%%%%%%%%%%%%%%%%%%%%%%%

\section{Newly Reported BBH mergers}
\label{sec:events}

Table~\ref{tab:signalsFoundParsPESourceFrame} lists some properties of the newly reported events: source-frame masses ($m_{1}^{\rm{src}}$ and $m_{2}^{\rm{src}}$), effective spin ($\chi_{\rm{eff}}$), redshift ($z$), inverse false alarm rate (IFAR) and estimated $p_{\rm astro}$. 

%Appendix~\ref{Appendix:posteriors} contains intrinsic parameter and redshift posteriors for all the new events, and PE samples are publicly available at \url{https://github.com/seth-olsen/new_BBH_mergers_O3a_IAS_pipeline}. 

We perform parameter estimation (PE) using a prior that is uniform in detector-frame constituent masses, effective spin $\chi_{\rm eff}$, and comoving volume-time ($VT$). The other extrinsic parameters have the standard geometric priors used in GWTC-3 \cite{lvc_gwtc3_o3_ab_catalog_2021}. We compute redshifts using a $\Lambda$ cold dark matter ($\Lambda$CDM) cosmology with Planck15 parameters \cite{cosmology_planck2015}. We use the phenomenological waveform approximant \texttt{IMRPhenomXPHM} \cite{xphm_pratten2020}, which includes the sub-dominant (or higher-order) modes $21$, $33$ and $44$ in addition to the dominant mode $22$, and models the effect of orbital precession due to in-plane spins. We accelerate the likelihood evaluations using the relative binning algorithm described in \citet{relative_binning} as implemented in the PE package $\tt{cogwheel}$ \cite{cogwheel} (see Appendix \ref{Appendix:posteriors} for more details). We note that these details are the same as \sk{what were used} in our O3a search: \citet{ias_o3a_pipeline2022}. 

In the rest of this section, we will briefly comment on the properties of the new events reported in this paper.

\subsection{High-mass sources}

\paragraph{\textbf{GW200109\_195634}}
This event is our most significant new detection, with $p_{\rm astro} = 0.81$. It has a primary mass ($m_{1}^{\mathrm{src}}=69^{+24}_{-19}\,\msun$) that lies in the UMG (see Fig.~\ref{fig:GW200109_195634_corner_plot} below). This event has a redshift ($z$) greater than 0.6 at $90\%$ confidence. 

We find that the best-fit waveform that the PE explores has a higher value of $\rm{SNR}^{2}$ than the search by  $\Delta \rm{SNR}^2 = 2\ln\mcal{L}_{\rm{max}} - \rho_{\rm{H}}^2 - \rho_{\rm{L}}^2 \approx 20$, where $\ln\mcal{L}_{\rm{max}}$ is the maximum log likelihood achieved within the posterior samples, and $\rho_{\rm{H, L}}$ represent the SNR recovered by the search template in the Hanford and Livingston detectors. It turns out that a significant fraction of the $\rm{SNR}^2$ difference ($\Delta \rm{SNR}^2 \approx 15$) comes from the Virgo detector which is included in the PE but not in our search. Including Virgo data in the future coincident searches could thus raise the significance of this event.

\paragraph{\textbf{GW200225\_075134}}
This event, with $\pastro = 0.60$, has a primary mass measurement of $m_{1}^{\mathrm{src}}=51^{+17}_{-11}\,\msun$ (Fig.~\ref{fig:GW200225_075134_corner_plot}). It does not pose an issue for the UMG.

\paragraph{\textbf{GW191117\_023843}} This event, with $\pastro = 0.56$, has a primary mass $m_{1}\gtrsim 47\, \msun$ at $90\%$ confidence. Thus,  this object could also be a BH in the UMG (Fig.~\ref{fig:GW191117_023843_corner_plot}).

\paragraph{\textbf{GW200210\_100022}} This event is one of the least significant events in our list, with $\pastro = 0.52$. The PE analyses confidently place the primary component in the UMG, with a measurement of $m_{1}^{\mathrm{src}} > 50 \,\msun$ at $90\%$ confidence (Fig.~\ref{fig:GW200210_100022_corner_plot}). The effective spin ($\chi_{\rm{eff}}$), on the other hand, is measured to be positive with a magnitude greater than $0.5$ at $90\%$ confidence, irrespective of the spin prior used. The mass ratio of the best-fit waveform is also high with $1/q \sim 4$. 

It might be worth noting that the PE analyses yield a slightly larger $\rm{SNR}^2$ than the search; $\Delta \rm{SNR}^2 = 2\ln\mcal{L}_{\rm{max}} - \rho_{\rm{H}}^2 - \rho_{\rm{L}}^2 \approx 8$. The additional $\rm{SNR}^2$ contribution comes from the HM and precession. This might be expected as the signal has a high mass ratio as well as a high total mass and effective spin. Recall that our search templates are based on the $\tt{IMRPhenomD}$ approximant \cite{IMRPhenomD_approximant_Khan2016}, which only models the dominant $(\ell, |m|) = (2, 2)$ mode and does not incorporate precession. Thus, a future search with templates that incorporate these physical effects may further improve the significance of this event.

\subsection{Low-mass sources}
\label{low_mass_sources}
\paragraph{\textbf{GW191228\_085854}}
\label{event_GW191228}

This event, with $p_{\rm astro} = 0.67$, has a bi-modal posterior distribution in the masses and effective spin under our default spin prior (see Fig.~\ref{fig:GW191228_085854_corner_plot}). The first posterior mode occurs at the secondary mass, $m_{2}^{\mathrm{src}} \sim 5.5\,\msun$ and effective spin, $\chi_{\rm{eff}} \sim -0.89$ while the second one occurs at $m_{2}^{\mathrm{src}} \sim 3 \,\msun$ and $\chi_{\rm{eff}} \sim  0.0$. Interestingly, the likelihoods of these two modes are very similar, which suggests that the data seems to prefer both modes equally. Both these solutions are, however, very interesting astrophysically. While the first solution hints at a highly negatively spinning stellar mass BBH merger, the second solution suggests at a significantly asymmetric ($1/q \gtrsim 5 $) non-spining BBH merger with the secondary mass confidently lying in the LMG. 

On the other hand, a PE with LVK prior (i.e., isotropic prior on spins) selects out only the second solution. This might be expected as this prior does not have significant weight at high positive or negative effective spins. In appendix ~\ref{Appendix:bimodality} we discuss this event in further detail and also show similar results (i.e., bi-modal phenomena) from known LVK events in the similar parameter space region when analysed with our default spin prior. This brings out to our attention the usefulness of using a generic prior so that at least the results for other priors of choice can be obtained by properly reweighting the generic result. 

% This event, with $p_{\rm astro} = 0.65$, has a secondary mass of $m_{2}^{\mathrm{src}}=5.5^{+0.9}_{-1.4}\,\msun$, near the top of the LMG. It also has a large negative effective spin, $\chi_{\rm{eff}} = -0.89^{+0.42}_{-0.10}$ . These properties, however, change with isotropic spin prior. The effective spin becomes consistent with zero, $\chi_{\rm{eff}} = -0.02^{+0.11}_{-0.22}$, while the mass ratio becomes quite asymmetric, $q = 0.14^{+0.06}_{-0.02}$, causing the secondary mass to fall confidently in the LMG, $m_{2}^{\mathrm{src}}=3^{+0.5}_{-0.2}\,\msun$. Such a change could be expected due to the well-known $q$--$\chi_{\rm{eff}}$ degeneracy present in the low total mass binaries (see e.g., \cite{Bai13}).

\begin{figure*}
    \centering
    \includegraphics[width=\linewidth]{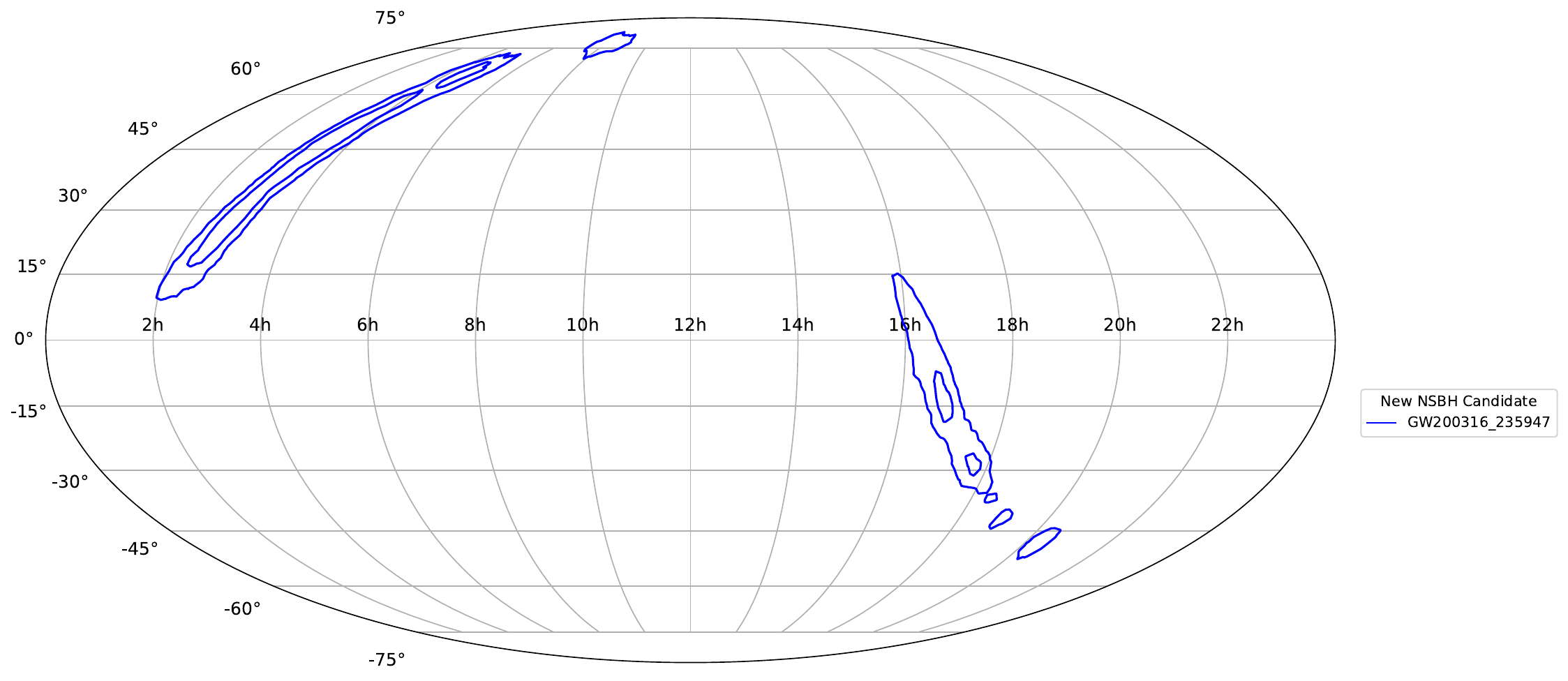}
    \vspace{-0.2cm}
    \caption{Sky localization for a new possible NSBH candidate, GW200316\_235947. The priors used were uniform in detector-frame constituent masses, effective spin, and comoving $VT$. The probability contours enclose $50\%$ and $90\%$. The $x$-axis represents right ascension in hours, and the $y$-axis represents declination in degrees.}
    \label{fig:GW200316_235947_sky_loc}
\end{figure*}

\paragraph{\textbf{GW200316\_235947}}:
This event is the most marginal event, with $p_{\rm astro} = 0.52$. It has a secondary mass of $m_{2}^{\mathrm{src}}=3.6^{+1.0}_{-1.3}\,\msun$ (Fig.~\ref{fig:GW200316_235947_corner_plot}), placing its lighter constituent confidently in the LMG. This object is also consistent with being a heavy NS at $90\%$ confidence, if we take the maximum NS mass to be $\sim 2.6\,\msun$. Based on the likelihood comparison between the regions $m_{2}^{\rm{src}}\lesssim 2.6\,\msun$ (NSBH) and $m_{2}^{\rm{src}}\gtrsim 2.6\,\msun$ (BBH), the NSBH and BBH solutions have comparable support from the data, so the priors may play a deciding role in identifying the secondary object as a BH or NS. The NSBH solution also has a slightly positive effective spin (see Fig.~\ref{fig:GW200316_235947_corner_plot}), which makes it more interesting for EM followup. In Fig.~\ref{fig:GW200316_235947_sky_loc} we show the sky localization of this event. Although the sky localization is not very well constrained, it would be worthwhile to do a follow-up search for an EM counterpart.% in the catalog of EM data at the given time and direction.  

%%%%%%%%%%%%%%%%%%%%%%%%%%%%%%%%%%%%%%%%%%%%%%  LVC EVENTS
%%%%%%

%NITZ
%Event GPS Time Observing Triggered astro IFAR (yr) ρH ρL 
%GW191224_043228 1261197166.15 HLV HL 0.87 0.13 5.0 6.8
%GW200106_134123 1262353301.93 HLV HL 0.69 0.06 5.2 5.2
%GW200129_114245 1264333383.11 HLV HL 0.53 0.04 5.1 6.0
%GW200210_005122 1265331100.74 HLV HL 0.74 0.04 5.4 6.3
%GW200214_223307 1265754805.00 HLV HL 0.72 0.08 5.2 5.2 
%GW200305_084739 1267433277.08 HLV HL 0.59 0.02 4.5 6.1
%GW200318_191337 1268594035.14 HLV HL 0.97 0.50 4.8 6.2
%%%%%%
%%%%%%
%%%
%%%

\section{Comparison to previously reported catalogs}
\label{sec:previous_events}

In Table~\ref{tab:lvc events}, we summarize our pipeline's results for the O3b Hanford--Livingston coincident events published in the GWTC-3 catalog \cite{lvc_gwtc3_o3_ab_catalog_2021}. We also include the additional events as well as the significance reported by the 4-OGC catalog \cite{nitz_4ogc_o3_ab_catalog_2021}. Our $\pastro$ estimation is based on the distribution of foreground and background triggers over the O3b run (see Appendix~\ref{Appendix:pastro}). We compute the inverse false-alarm rate independently for each template bank. For a comparison of $\pastro$ across different search pipelines, see Section IV of \citet{ias_o3a_pipeline2022}. 
%\seth{[should we mention that the background is estimated with time slides? we should mention that somewhere here, but maybe it is later]}\ajitm{[Ajit: I have mentioned this in Appendix \ref{Appendix:pastro}, where we discuss the estimation of $\pastro$.]}

%some notes:
%2 light events we missed have Virgo but there is no SNR breakdown by detector in the GWTC-3 paper, so not sure yet if our failure was just missing Virgo or something else:
%-- GW200115\_042309 was a failure of BBH\_0 but maybe need NSBH templates (they get m1, m2 = 5.8, 1.5), found by PyCBC-Broad, GstLAL, and MBTA, plus nitz
%-- GW200202\_154313 was a failure of BBH\_1, found by PyCBC-BBH and GstLAL, plus nitz
%low SNR for ones in BBH\_3 that nitz also missed, and several vetoes in BBH\_4 and BBH\_5 (some missed by nitz)

\paragraph{\textbf{Confidently recovered events}} The LVK collaboration has five main search pipelines: cWB \cite{Klimenko:2016},  MBTA \cite{Andres:2021vew}, GstLAL \cite{Sachdev:2019vvd}, PyCBC and PyCBC\_BBH \cite{Usman:2015kfa}. Whenever the estimated $\pastro$ from any one of these pipelines crosses the threshold of one half, the event is declared a confident detection. With this criterion, there were 32 Hanford--Livingston (HL) coincident events in \citet{lvc_gwtc3_o3_ab_catalog_2021}. However, two of them (GW200115\_042309 and GW191219\_163120) were NSBH mergers. We did not look for such signals in this work. Of the remaining 30 coincident BBH mergers, our pipeline retained \neviasconf{} with a significance comparable to or better than the LVK catalog. 
%We also recovered two of the additional seven events reported in 4-OGC catalog: GW191224\_043228 and GW200129\_114245.
% Eight LVK events were marginally recovered, i.e., with a significance below one-half. Four events were vetoed (indicated by the word
% “Veto” in the IFAR column) while two events were missed.  We discuss these below.

\paragraph{\textbf{Candidates with low $\pastro$}}
Our pipeline assigned $\pastro < 0.5$ to 6 of the LVK events. Five of these events were detected confidently by only one of the LVK search pipelines: GW191126\_115259, GW200208\_222617 and GW200308\_173609 were detected by only PyCBC\_BBH, 
while GW200306\_093714 and GW200322\_091133 were detected by only MBTA. Given that each individual pipeline differs in their analysis choices, which can be especially important for near-threshold events like these, it is not surprising that these signals were also considered marginal by our pipeline. The remaining event, GW200202\_154313, was detected by more than one pipeline. However, this event also has a substantial SNR contribution from the Virgo detector, which indicates that adding Virgo data to our coincident search may improve its significance in our future analyses.

%GW191103\_012549 was just bad. BBH\_2 (12+8 msun at 1 Gpc)
%GW191204\_110529 was a BBH\_4 or BBH\_5 candidate (27+19 msun at 1.8 Gpc)
%-- MAYBE should have been BBH\_4 but got vetoed so kept BBH\_5 trigger.
%GW200115\_042309 is a BBH\_0 or (1, 0) prospective NSBH signal (9+1.9 msun at .55 Gpc)
%GW200202\_154313 is a BBH\_1 candidate (10+7 msun at .41 Gpc)
%GW200208\_222617 is a BBH\_3 candidate (51+12 msun at 4 Gpc) 
%GW200306\_093714 (28+15 msun at .4 Gpc) was an event which only MBTA recovered with $\pastro > .5$, whereas no matched filter pipelines crossed the detection threshold.
%GW200308\_173609 is a BBH\_3 candidate (36+14 msun at 5.4 Gpc) which only PyCBC-BBH recovered with $\pastro > .5$, whereas no other pipelines crossed the detection threshold (with MBTA at $\pastro = .24$ and the others below 1\%).
%GW200322\_091133 (34+14 msun at 3.6 Gpc) was an event which only MBTA recovered at $\pastro > .5$, with both PyCBC pipelines finding less than 10\%.

\paragraph{\textbf{Vetoed candidates}}

We implement a series of signal consistency checks in order to reject triggers that are likely to arise from glitches (noise transients) and consequently improve our pipeline's sensitivity. 
Noise fluctuations and the incompleteness of our template bank can cause astrophysical signals to be vetoed, but the tests are designed to gain more in true positive rate than the loss they produce from false negatives due to these effects (in a statistical sense). 

Six events from the GWTC-3 catalog were vetoed in our search: GW191103\_012549,  GW191109\_010717, GW191113\_071753, GW191127\_050227, GW200129\_065458 and GW200210\_092254. 
 
The events GW191109\_010717 and GW200129\_065458 had fairly high SNR and are high-mass binaries with imprints of precession \cite{lvc_gwtc3_o3_ab_catalog_2021}. Our search used a template bank that was built using the non-precessing $\tt{IMRPhenomD}$ approximant, and the signal consistency tests compare the data only to the best-fit waveform within the template bank -- the absence of precession and higher-harmonics in the template can make it more likely for events that exhibit these phenomena to trigger the signal-consistency tests~\footnote{We note that some of the vetoed events here are restored in our search with higher-harmonics~\cite{Wad23_Events}.}. The data for two events was also flagged for data-quality issues in the GWTC-3 catalog. {Currently, the triggers that fail any of our $\sim 45$ separate veto tests are completely excluded from our catalog. In the future, we plan to devise a scheme for including the failed-test candidates back into our catalog, but with an appropriate penalty. This approach would ensure that high-significance events are not entirely removed due to the absence of certain physical effects in our templates, though this may come at the cost of slightly increasing the background in our search.}

Our pipeline vetoed GW191113\_071753 since the residuals after subtracting the best-fit template exhibited excess power beyond the expectation from Gaussian noise in multiple frequency bands. 
In the GWTC-3 analysis, this candidate was marked as affected by transient, non-Gaussian detector noise. 
The best-fit waveform in PE for this candidate has a significantly asymmetric mass ratio $q\lesssim 0.2$, and the absence of higher modes in our template bank could have also contributed to the failure of the excess power tests. 

GW200210\_092254 failed ``split-tests" designed to check for consistency between the SNR in different sub-ranges of frequency. 
This event also has a significantly asymmetric mass ratio, but when we investigated further, we found that our pipeline assigned a representative trigger to this event that came from a different region of the template bank than would be implied from the parameters from the inference runs. 
The template with the correct parameters had a matched-filtering SNR in Hanford that fluctuated below the collection threshold we used to define triggers\footnote{We set these thresholds to reduce the large volume of trigger data that we would otherwise need to analyze with the more costly coincidence-section of the pipeline.}. 
In other words, the actual trigger in the Livingston data could not find a coincident one in the Hanford data due to this threshold, and thus the event was assigned to a different bank in which it failed signal consistency tests. 
We can detect such events by either lowering the SNR threshold for triggering, or performing a targeted search for events with disparate detector-responses, such as in Ref.~\cite{fishingIAS19_prd2021}.

The other two events were straightforwardly vetoed in our search; GW191127\_050227 failed multiple excess power tests in a few specific bands in a unit-second interval, while GW191103\_012549 failed the split tests. We also note that the data for GW191127\_050227 was flagged as being affected by data-quality issues in the GWTC-3 catalog paper.
%This event was detected with a very high SNR ($21.2$) in L1 detector by Gst-LAL. We also see that the veto tests for this event were in L1. 

We further note that, two events GW191129\_134029 and GW200202\_154313 were recovered with secondary triggers, i.e., their primary triggers were vetoed. This is the reason we recover them with lower SNR, and thus the lower IFAR compared to the LVK analysis (see Table~\ref{tab:lvc events}).

% GW191109\_010717 was a BBH\_5 signal (65+47 msun at 1.3 Gpc) that was confidently recovered by all LVC pipelines.
% GW191113\_071753 was a BBH\_2 candidate which only MBTA recovered with $\pastro > .5$, whereas no matched filter pipelines crossed the detection threshold. GLITCH?
% GW200129\_065458 was a candidate near a hole, with odd behavior in Livingston, and may have had a better chance with a different subbank.
% GW200210\_092254 is a BBH\_1 candidate (24+2.8 msun at 1 Gpc)

\paragraph{\textbf{Lost triggers}} Our pipeline missed the event GW200220\_061928. This is a high-mass BBH (87\,$\msun+61\,\msun$ at 6 Gpc) which was only found by PyCBC-Broad, with all other LVK pipelines having no coincident triggers at this time. 
%This event was also a marginal trigger in 4-OGC \cite{nitz_4ogc_o3_ab_catalog_2021}.

%GW191219\_163120 is a BBH\_0 or (1, 0) prospective NSBH signal (31+1.2 msun at .27 Gpc) which had no coincident candidates in our search. Only PyCBC-BBH found this, with all other LVC pipelines having no coincident triggers at this time.

%GW200220\_061928 is a BBH\_5 signal (87+61 msun at 6 Gpc) which had no coincident candidates in our search. Within a second of this time there was a trigger with a squared SNR of roughly 44 in Livingston for bank (5, 5) but nothing above 24 in Hanford, and nothing coherent. Only PyCBC-Broad found this, with all other LVC pipelines having no coincident triggers at this time.

\paragraph{\textbf{Search space excluded from this work}}

We did not perform a dedicated search for NSBH or BNS signals. We leave this to future work.

\bgroup
\def\arraystretch{1.25}
\begin{table*}
    \centering
    \caption{Hanford--Livingston coincident BBH events already reported in the GWTC-3 catalog \cite{lvc_gwtc3_o3_ab_catalog_2021} as detected by our pipeline and the 4-OGC catalog \cite{nitz_4ogc_o3_ab_catalog_2021}. Six events found by LVK in Hanford--Livingston coincidence were vetoed in our search: GW191103\_012549,  GW191109\_010717, GW191113\_071753, GW191127\_050227, GW200129\_065458 and GW200210\_092254. The
inverse false alarm rate (IFAR) values in the GWTC-3 column are taken from the \href{https://gwosc.org/eventapi/json/GWTC-3-confident/}{$\tt{GWOSC}$} catalog, which
corresponds to whichever LVK pipeline achieved the highest astrophysical probability for that event in the GWTC-3
analysis. The 4-OGC column was taken from the catalog summary data at \href{https://github.com/gwastro/4-ogc/blob/master/search/4OGC_top.txt}{$\tt{GitHub}$}, as listed in the publication.
    The events in {bold} are the ones from the 4-OGC catalog \cite{nitz_4ogc_o3_ab_catalog_2021}.}
    \begin{tabular}{|c|c|cc|c|ccc|ccc|}%cccc|}
        %\hline
        \hline
        \multirow{2}{*}{Event Name} & \multirow{2}{*}{Bank} & \multirow{2}{*}{$\rho_{\rm H}$} & \multirow{2}{*}{$\rho_{\rm L}$} & \multirow{2}{*}{${}^{\rm GWTC-3}$}  & \multicolumn{3}{c|}{$p_{\rm astro}$} & \multicolumn{3}{c|}{IFAR (yr)} \\
        %& \multirow{2}{*}{$m_1 {\scriptstyle( \rm M_\odot)}$} & \multirow{2}{*}{$m_2 {\scriptstyle(\rm M_\odot)}$} & \multirow{2}{*}{$\chi_{\rm eff}$} & \multirow{2}{*}{$z$}
         \cline{6-11}
         & & & & $\rho_{\rm Network}$\footnote{Note that the quoted {$\rm{SNR}$} is from the parameter estimation runs for the events [\href{https://gwosc.org/eventapi/json/GWTC-3-confident/}{$\tt{GWOSC}$}]. The search {$\rm{SNR}$} is generally lower and varies between the different pipelines used by the LVK.} & \, IAS \, $|$ & $|$ GWTC-3 $|$ & $|$ 4-OGC & \,
         IAS~\footnote{The combined IFAR across the chirp-mass banks.} \, $|$ & $|$ GWTC-3 $|$ & $|$ 4-OGC \\
         %& & & & \\
         \hline
\rowcolor{gray!30} GW191103\_012549 & \texttt{BBH\_2} & $6.3$ & $6.6$ & $8.9$ & $ -- $ & $0.94$ & $--$ & Veto & $2.2$ & $--$ \\
GW191105\_143521 & \texttt{BBH\_1} & $5.6$ & $7.6$ & $9.7$ & $0.99$ & $0.99$ & $1.0$ & $> 1000$ & $83.3$ & $316$ \\
\rowcolor{gray!30} GW191109\_010717 & \texttt{BBH\_5} & $9.2$ & $12.9$ & $17.3$ & $ -- $ & $0.99$ & $1.0$ & Veto & $> 1000$ & $> 1000$ \\
GW191113\_071753 & \texttt{BBH\_2} & $6.1$ & $5.4$ & $7.9$ & $ -- $ & $0.68$ & $--$ & Veto & $0.038$ & $--$ \\
\rowcolor{gray!30} GW191126\_115259 & \texttt{BBH\_2} & $5.8$ & $6.6$ & $8.3$ & $0.36$ & $0.70$ & $1.0$ & $0.20$ & $0.31$ & $4.9$ \\
GW191127\_050227 & \texttt{BBH\_2} & $7.0$ & $6.5$ & $9.2$ & $ -- $ & $0.74$ & $0.99$ & Veto & $4.0$ & $0.15$ \\
\rowcolor{gray!30} GW191129\_134029\footnote{\label{primaryvetoed} The primary triggers for these event were vetoed. We recover them with secondary triggers with much lower SNRs, which can happen rarely due to parameter choices in the clustering of single-detector triggers in the pipeline.} & \texttt{BBH\_1} & $5.3$ & $7.1$ & $13.1$ & $0.69$ & $0.99$ & $1.0$ & $2.2$ & $> 1000$ & $> 1000$ \\
GW191204\_110529 & \texttt{BBH\_5} & $4.6$ & $7.0$ & $8.9$ & $0.68$ & $0.74$ & $0.99$ & $1.1$ & $0.30$ & $1.6$ \\
\rowcolor{gray!30} GW191204\_171526 & \texttt{BBH\_1} & $9.4$ & $13.5$ & $17.4$ & $1.00$ & $0.99$ & $1.0$ & $> 1000$ & $> 1000$ & $> 1000$ \\
GW191215\_223052 & \texttt{BBH\_3} & $7.2$ & $8.1$ & $11.2$ & $1.00$ & $0.99$ & $1.0$ & $> 1000$ & $> 1000$ & $869$ \\
\rowcolor{gray!30} GW191219\_163120 & -- & $ -- $ & $ -- $ & $9.1$ & $ -- $ & $0.82$ & $--$ & $ -- $ & $0.25$ & $--$ \\
GW191222\_033537 & \texttt{BBH\_4} & $8.6$ & $8.2$ & $12.5$ & $1.00$ & $0.99$ & $1.0$ & $> 1000$ & $> 1000$ & $> 1000$ \\
\rowcolor{gray!30} \textbf{GW191224\_043228} & \texttt{BBH\_2} & $5.4$ & $7.1$ & $--$ & $0.45$ & $--$ & $0.87$ & $0.36$ & $--$ & $0.13$ \\
GW191230\_180458 & \texttt{BBH\_4} & $7.3$ & $6.7$ & $10.4$ & $1.00$ & $0.96$ & $1.0$ & $> 1000$ & $20.0$ & $497$ \\
\rowcolor{gray!30} \textbf{GW200106\_134123} & \texttt{BBH\_4} & $5.5$ & $5.4$ & $--$ & $0.40$ & $--$ & $0.69$ & $0.26$ & $--$ & $0.059$ \\
GW200115\_042309 & \texttt{BBH\_0} & $5.4$ & $7.6$ & $11.3$ & $0.12$ & $0.99$ & $-1.0$ & $0.018$ & $> 1000$ & $940$ \\
\rowcolor{gray!30} GW200128\_022011 & \texttt{BBH\_4} & $7.3$ & $6.8$ & $10.6$ & $1.00$ & $0.99$ & $1.0$ & $> 1000$ & $233$ & $307$ \\
GW200129\_065458 & \texttt{BBH\_4} & $13.6$ & $18.3$ & $26.8$ & $ -- $ & $0.99$ & $1.0$ & Veto & $> 1000$ & $> 1000$ \\
\rowcolor{gray!30} \textbf{GW200129\_114245} & \texttt{BBH\_4} & $4.7$ & $6.5$ & $--$ & $0.38$ & $--$ & $0.53$ & $0.22$ & $--$ & $0.037$ \\
GW200202\_154313$^{\ref{primaryvetoed}}$ & \texttt{BBH\_1} & $5.3$ & $6.2$ & $10.8$ & $0.008$ & $0.99$ & $1.0$ & $0.003$ & $> 1000$ & $6.1$ \\
\rowcolor{gray!30} GW200208\_130117 & \texttt{BBH\_3} & $6.7$ & $7.2$ & $10.8$ & $1.00$ & $0.99$ & $1.0$ & $> 1000$ & $> 1000$ & $917$ \\
GW200208\_222617 & \texttt{BBH\_3} & $5.1$ & $5.8$ & $7.4$ & $0.14$ & $0.70$ & $--$ & $0.029$ & $0.21$ & $--$ \\
\rowcolor{gray!30} GW200209\_085452 & \texttt{BBH\_3} & $7.3$ & $5.9$ & $9.6$ & $0.95$ & $0.97$ & $0.99$ & $44.0$ & $21.7$ & $1.1$ \\
\textbf{GW200210\_005122} &\texttt{BBH\_1} & $5.5$ & $5.9$ & $--$ & $--$ & $--$ & $0.74$ & Veto & $--$ & $0.042$ \\
\rowcolor{gray!30} GW200210\_092254 & \texttt{BBH\_1} & $5.0$ & $6.5$ & $8.4$ & $ -- $ & $0.54$ & $--$ & Veto & $0.83$ & $--$ \\
\textbf{GW200214\_223306} & \texttt{BBH\_4} & $5.5$ & $5.1$ & $--$ & $0.23$ & $--$ & $0.72$ & $0.069$ & $--$ & $--$ \\
\rowcolor{gray!30} GW200216\_220804 & \texttt{BBH\_4} & $6.6$ & $5.4$ & $8.1$ & $0.62$ & $0.77$ & $0.78$ & $1.2$ & $2.9$ & $0.093$ \\
GW200219\_094415 & \texttt{BBH\_3} & $6.1$ & $8.2$ & $10.7$ & $1.00$ & $0.99$ & $1.0$ & $> 1000$ & $> 1000$ & $22.9$ \\
\rowcolor{gray!30} GW200220\_061928 & -- & $ -- $ & $ -- $ & $7.2$ & $ -- $ & $0.62$ & $--$ & $ -- $ & $0.15$ & $--$ \\
GW200220\_124850 & \texttt{BBH\_4} & $6.2$ & $5.3$ & $8.5$ & $0.98$ & $0.83$ & $--$ & $264$ & $0.033$ & $--$ \\
\rowcolor{gray!30} GW200224\_222234 & \texttt{BBH\_4} & $12.5$ & $13.0$ & $20.0$ & $1.00$ & $0.99$ & $1.0$ & $> 1000$ & $> 1000$ & $> 1000$ \\
GW200225\_060421 & \texttt{BBH\_2} & $9.5$ & $7.8$ & $12.5$ & $1.00$ & $0.99$ & $1.0$ & $> 1000$ & $> 1000$ & $> 1000$ \\
\rowcolor{gray!30} \textbf{GW200305\_084739} & \texttt{BBH\_3} & $4.2$ & $6.7$ & $--$ & $0.31$ & $--$ & $0.59$ & $0.14$ & $--$ & $0.019$ \\
GW200306\_093714 & \texttt{BBH\_3} & $5.6$ & $5.9$ & $7.8$ & $0.26$ & $0.81$ & $0.51$ & $0.098$ & $0.042$ & $0.018$ \\
\rowcolor{gray!30} GW200308\_173609 & \texttt{BBH\_3} & $4.8$ & $6.2$ & $4.7$ & $0.16$ & $0.86$ & $--$ & $0.037$ & $0.42$ & $--$ \\
GW200311\_115853 & \texttt{BBH\_3} & $12.5$ & $10.3$ & $17.8$ & $1.00$ & $0.99$ & $1.0$ & $> 1000$ & $> 1000$ & $817$ \\
\rowcolor{gray!30} GW200316\_215756 & \texttt{BBH\_2} & $5.5$ & $8.1$ & $10.3$ & $0.92$ & $0.99$ & $1.0$ & $29.3$ & $> 1000$ & $22.4$ \\
\textbf{GW200318\_191337} &\texttt{BBH\_4} & $4.73$ & $6.5$ & $--$ & $--$ & $--$ & $0.97$ & Veto & $--$ & $0.50$ \\
\rowcolor{gray!30} GW200322\_091133 & \texttt{BBH\_3} & $5.6$ & $5.4$ & $4.5$ & $0.051$ & $0.62$ & $--$ & $0.008$ & $0.007$ & $--$ \\

         \hline
         %\hline
    \end{tabular}
    \label{tab:lvc events}
\end{table*}
\egroup

%%%%%%%%%%%%%%%%%%%%%%%%%%%%%%%%%%%%%%%%%%%%%%%%%%%%%%%%%
%%%%%%%%%%%%%%%%%%%%%%%%%%%%%%%%%%%%%%%%%%%%%%%%%%%%%%%%%
%%%%%%%%%%%%%%%%%%%%%%%%%%%%%%%%%%%%%%%%%%%%%%%%%%%%%%%%%
%%%%%%%%%%%%%%%%%%%%%%%%%%%%%%%%%%%%%%%%%%%%%%%%%%%%%%%%%
\section{Discussion}
\label{sec:conclusions}

Below we present a qualitative discussion of the astrophysical properties of the new events, including where they could possibly contribute to constraints from full population analyses. We plan to conduct a population analysis involving these events in a separate work. We conclude with the summary of our O3b search results and an outline of the major improvements for our upcoming searches unifying O3a and O3b data. 

\subsection{Astrophysical implications of the new events} \label{sec:astro_implications}
\paragraph{\textbf{The lower mass gap (LMG)}}
The BH masses inferred from observations of low-mass X-ray binaries \cite{Bailyn_1998,Ozel_2010,Farr_2011} seem to be concentrated above $\sim 5\,\msun$. Separately, there are observational upper limits on the known NS mass range ($\lesssim  2.6\,\msun$) \cite{max_NS_mass_est_from_EOS_space_alsing2018berti}. Thus, there might be no way to form compact objects between $\sim 2.6\,\msun$ and $\sim 5\,\msun$. This is the basis for hypothesizing the LMG in the astrophysical mass distribution. However, it is still unclear whether the existence of such a gap is supported by any solid theoretical or observational evidence (as opposed to, for example, an artifact of systematic effects in the EM observations of x-ray binaries and GW observations of BBH, NSBH, and BNS mergers). 

As of now, we have already observed GW sources whose secondary constituents confidently lie below $\sim 5\,\msun$ \cite{ias_o3a_pipeline2022, lvc_gwtc3_o3_ab_catalog_2021}. Population analyses using catalogs that include the LVK O3b events \cite{lvc_o3a_population_properties_2021, ias_o3a_population_analysis_prd2021roulet, LVKO3bpopulation} also do not provide strong evidence for such a gap. Although a substantial reduction in the merger rates above the maximum NS mass ($\sim 2.5\,\msun$) is observed, there is no subsequent rise above $\sim 5\,\msun$ (see figure~5 of \cite{LVKO3bpopulation}). This indicates that two distinct populations are required for modeling the observed compact objects, but otherwise no such gap is discernible. The detection of more events in this region of parameter space will help us gain a stronger understanding of this question.

Here we present a new event, GW200316\_235947, whose secondary constituent might be either a BH in the LMG or a heavy NS. The NS scenario may be more interesting from an astrophysical standpoint.  A prior that includes the LMG as a feature would indeed yield only the NSBH solution. This solution would also have a positive effective spin and mass ratio $1/q \gtrsim 3.5$. The positive spin makes it more favorable for an EM counterpart follow-up search, and we show the sky localization for this event in Fig.~\ref{fig:GW200316_235947_sky_loc}. The BBH solution, $m_{2}^{\rm{src}} \gtrsim 2.6\,\msun$, has a slightly negative effective spin with nearly equal mass ratio. Thus, this event looks interesting under either of the potential solutions for the source configuration. Since it has a low $\pastro$, its impact on the population analysis may be limited \cite{ias_popO2_Roulet_2020}. However, the population analysis also recomputes $\pastro$ under the various astrophysical priors being tested, so there is a chance that it would become more important in the case of other mass and spin priors.

Another event, GW191228\_085854, might have its secondary constituent in the LMG. We saw that, under our default (uniform) prior on the effective spin, this event yielded bi-modal posterior distribution on the secondary mass where the secondary mode lies confidently in the LMG. This solution was also preferred by the isotropic spin prior (LVK prior). This solution also has a high mass ratio, $1/q \gtrsim 5$ at $90\%$ confidence interval.

% this event is a BBH above the LMG with its constituent spins strongly anti-aligned ($\chi_{\rm{eff}} = -0.89^{+0.42}_{-0.10}$). However, under the isotropic spin prior, the preferred source is non-spinning with significantly asymmetric masses ($1/q \gtrsim 5$) and a secondary sitting confidently in the LMG. Based on the likelihood and Bayesian evidence comparisons, we find that the former solution is slightly favored (by a Bayes factor $\sim e^{1.5}$). Nevertheless, each solution has atypical properties, making this event interesting to study in more detail.

\paragraph{\textbf{The upper mass gap (UMG)}}
{The evolution of massive stars can involve phases such as PPISN and PISN, where the stars either lose a significant fraction of their mass or are completely disrupted, leaving no remnant. As a result, PISN create a gap in the BH mass distribution, while PPISN can populate the mass spectrum below this gap. The precise boundaries of this gap remain uncertain, as they depend on several factors, including the rate of carbon-to-oxygen burning in the core (${}^{12}\rm{C}(\alpha, \gamma){}^{16} O$), angular momentum transport, metallicity, and wind loss. \cite{Marchant:2020haw, Farmer:2019jed, Mehta:2021fgz}. }
%For example, the uncertainty in the nuclear reaction rate ${}^{12}\rm{C}(\alpha, \gamma){}^{16} O$ can move the lower edge of the mass gap to anywhere between $\sim 45\,\msun$ and $\sim 90\,\msun$ \footnote{While the upper edge changes from $\sim 90\,\msun$ to $\sim 170\,\msun$} (see Fig. 10 of \cite{Mehta:2021fgz}). 

Analyses thus far have not found any conclusive evidence for the UMG in the observed BH mass distribution \cite{LVKO3bpopulation, ias_o3a_population_analysis_prd2021roulet}. This either challenges the current theoretical models of stellar evolution, or the gap might be filled by BBH formed via alternative mechanisms such as hierarchical mergers \cite{hierarchical_7merger_scenario2020b, hierarchical_from_dynamical_in_any_star_cluster2020b, hierarchical_mergerFamily_dynamical_mass_dist_matters2021, hierarchical_rate_sensitive_to_natal_spins_Fragione2021kocsis}. Since the prediction of relative rates of mergers throughout parameter space is highly uncertain, it is hard to tell at this time. As the number of observations in this region of parameter space grows, the structures in the observed BH mass distribution should be able to tell us whether such a gap really exists. 

Here we present three new events, GW191117\_023843, GW200109\_195634 and GW200210\_100022, whose primary masses confidently lie between $\sim 45\,\msun$ and $\sim 135\,\msun$. Our measurement of the effective spin for the first two events is consistent with zero. However, for the third event (GW200210\_100022), the effective spin is measured to be greater than $0.5$ at $90\%$ confidence. The inference holds regardless of the spin prior used, suggesting this event could be a highly positively spinning BH binary. Even for the second event, GW200109\_195634,  although the Bayesian analysis yields an effective spin measurement consistent with zero,  the high likelihood region is concentrated at a positive effective spin ($\chi_{\rm{eff}}\sim 0.5$). The spin properties make these events stand out among the existing detections. The significance of these events will likely be improved in future searches where HM and precession are included in the templates and with Virgo data included in the coincident triggering. 

%Another event GW191121\_101441, which is our most secured event, has the primary mass sitting above $\sim 45\,\msun$. However, 

\paragraph{\textbf{Spin}}
GW signals also allow us to measure the spins of the binary constituents. Specifically, a certain combination of these spins (namely, the effective spin) can be measured more accurately \cite{AjithSpin,Ng:2018neg}. These spin parameters encode crucial information about BBH formation channels. For example, the isolated formation channel (under standard assumptions for parameters like BH natal kicks) is highly unlikely to form anti-aligned or misaligned BBH systems, especially if the total mass is high \cite{need_dynamical_to_misalign_orbit_and_spin_Rodriguez2016}. The characteristics of BBH spins depend on many physical processes that occur during the evolution of the progenitor stars, e.g., the angular momentum transport \cite{2019MNRAS.485.3661F, Fuller_2019, Belczynski2020}, tides and mass transfer \cite{PhysRevD.98.084036,Qin_2019, Bavera2020}, external environments, BH natal kicks \cite{need_dynamical_to_misalign_orbit_and_spin_Rodriguez2016}, etc. It is expected that BBH formed in isolated environments will have spins preferentially aligned with the orbital angular momentum. On the other hand, BBH formed in dense stellar environments (such as globular clusters) will have their spin directions randomized by interactions, i.e., an isotropic spin distribution \cite{2016ApJ...832L...2R, Mandel_2010}.

The GW detectors are slightly biased towards positively aligned ($\chi_{\rm{eff}}>0$) BBH systems due to selection effects.\footnote{This is because the $\chi_{\rm{eff}} > 0$ systems are relatively louder for fixed mass and distance.} This means that we can observe these systems from a larger volume of space. Thus, their fraction in the population of GW observations is expected to be correspondingly larger than their fraction of the true astrophysical population. This is consistent with what we see in Fig.~\ref{fig:population}. This selection effect is indeed taken into consideration during the population analysis, and thus the inference of astrophysical parameters is not expected to be biased. 

%However, it is not clear that \ajitm{the lower fraction of $\chi_{\rm{eff}} < 0$ systems} in Fig.~\ref{fig:population} is entirely due to selection effects. We know that the anti-aligned systems are shorter in length and hence they could be more prone to being mistaken for glitches. The waveform approximants in this regime might also not be very faithful. One way to address these questions would be to do injection studies where we inject astrophysical signals into the data and let the pipeline discover them. We plan to do this in our next work.

There are some interesting spin properties among the new events reported here which could be informative in the full population analysis. The high-mass binaries GW200210\_100022 and GW200109\_195634 are likely to have high positive effective spin ($\chi_{\rm{eff}}\gtrsim 0.5$). On the other hand,  the low-mass binary GW191228\_085854 may be a merger with a large negative $\chi_{\rm{eff}} = -0.89^{+0.42}_{-0.10}$. If true, this may add valuable information to the full population analysis as there is rarely such an event in this parameter space.

% However, its properties change under the isotropic spin prior, where it becomes a non-spinning (but significantly asymmetric-mass) signal due to the known $q$--$\chi_{\rm{eff}}$ degeneracy. The latter property may also \sk{be useful for}\seth{have some effect on} the population analysis. 

\subsection{Concluding remarks}

In this work, we report \neviasnew{} new candidate BBH mergers from the IAS pipeline's search of O3b Hanford--Livingston coincident data, based on the criteria that the astrophysical probability is greater than one half (following the LVK collaboration \cite{lvc_gwtc3_o3_ab_catalog_2021}). We recovered 17 of the 30 Hanford--Livingston coincident BBH events with comparable or improved significance. Our pipeline recovered 6 events with a significance lower than one half; most of these events were detected by only one LVK pipeline. Thus, in total, we present \neviastot{} confidently detected events in the Hanford--Livingston coincident data from the second part of LVK's third observing run (O3b). The new events reported here, except GW200225\_075134 and GW191117\_023843, are also present in the list of GWTC-3 sub-threshold events released by the LVK collaboration \cite{GWTC-3Marginal}.

The new events declared here have interesting astrophysical properties. For example, three events have their primary constituents sitting in the UMG, one of which might also be highly positively spinning with $\chi_{\rm{eff}}\geq 0.5$ at $90\%$ confidence. One event has its secondary constituent placed confidently below the top of the LMG, with a possibility of being an NSBH signal. We also have a low-mass binary which may have strongly anti-aligned spins. As we can see from Fig.~\ref{fig:population}, at least four events lie in the sparsely populated regions of the $\chi_{\rm{eff}}$--$M^{\rm{src}}_{\rm{tot}}$ plane. These \sk{findings suggest that the} new events could statistically impact the inferred distributions of BBH component masses and effective spin. 

We expect roughly \nevfalseposias{} of the newly reported events to be noise transients rather than astrophysical signals, as suggested by summing the complements of the $\pastro$ values in Table~\ref{tab:signalsFoundParsPESourceFrame}. Nevertheless, we have shown that for at least two events the $\pastro$ may be improved in future searches that include Virgo data in coincident analysis, and HM and precession in search templates. The source properties, including significance ($\pastro$), become more sensitive to the prior choices for low-SNR events like ours. To facilitate further analysis and comparisons between priors, we make our posterior samples and other relevant information public at \url{https://github.com/AjitMehta/new_BBH_mergers_O3b_IAS_pipeline}. These results can be reweighted with other choices of priors to observe how the results (like the number of events itself) may change. Our prior choices differ from the other catalogs \cite{lvc_gwtc3_o3_ab_catalog_2021, nitz_4ogc_o3_ab_catalog_2021} mainly in the spins. We use a prior that is flat in the effective spin. Our motivation lies in the fact that the effective spin is the best-measured spin parameter, and hence we want to give equal weights to all physically possible values of that parameter \cite{flat_chieff_prior_o1event1216_formation_channels2019}. On the other hand, using an isotropic spin prior will significantly down-weight large effective spins. A full population analysis not only requires the PE samples, but also the $\pastro$ of each event \cite{ias_popO2_Roulet_2020}.  For this reason, we also provide a file with the full list of triggers, \texttt{IAS\_O3b\_triggers.hdf}, in the Github repository. 

%\seth{\sout{All these quantities are thus needed to be recomputed as a function of the hyper-parameters of a given astrophysical model. Therefore, more events than reported in this work could enter the list of events being considered for the population analysis.}}

The idea of this work was to complete the search on the full third observing run (O3) data using the same pipeline that was used for our O3a search \cite{ias_o3a_pipeline2022}. Nonetheless, we already made a few notable improvements in our pipeline for this work, and we continue to plan to make improvements for upcoming searches. In fact, we \sk{are already doing} will soon release results from a search of the full O3 data using templates that include higher harmonics\sk{. The result should be published soon} \cite{Wad23_Events}. We also plan to include Virgo detector data in our coincident analysis in future work. 
Our next immediate task is to do an injection study, which should enable us to assess the sensitivity of our pipeline in different regions of the compact binary parameter space.

%%%%%%%%%%%%%%%%%%%%%%%%%%%%%%%%%%%%%%%%%%%%%%%%%%%%%%%%%%%%%%%%%%%%%%%%%%%%
\section*{Acknowledgements}
%%%%%%%%%%%%%%%%%%%%%%%%%%%%%%%%%%%%%%%%%%%%%%%%%%%%%%%%%%%%%%%%%%%%%%%%%%%%
% D. W. gratefully acknowledges support from the Friends of the Institute for Advanced Study Membership and from the W. M. Keck Foundation Fund. MZ is supported by NSF 2209991 and NSF-BSF 2207583.
% BZ is supported by NSF-BSF 2207583 and ISF

We thank Horng Sheng Chia, Katerina Chatziioannou for helpful discussions. 
DW gratefully acknowledges support from the Friends of the Institute for Advanced Study Membership and the Keck foundation. 
TV acknowledges support from NSF grants 2012086 and 2309360, the Alfred P. Sloan Foundation through grant number FG-2023-20470, the BSF through award number 2022136 and the Hellman Family Faculty Fellowship. BZ is supported by the Israel Science Foundation, NSF-BSF and by a research grant from the Willner Family Leadership Institute for the Weizmann Institute of Science. MZ is supported by NSF 2209991 and NSF-BSF 2207583. This research was also supported in part by the National Science Foundation under Grant No. NSF PHY-1748958. We also thank ICTS-TIFR for their hospitality during the completion of a part of this work. 

This research has made use of data, software and/or web tools obtained from the Gravitational Wave Open Science Center (\url{https://www.gw-openscience.org/}), a service of LIGO Laboratory, the LIGO Scientific Collaboration and the Virgo Collaboration. LIGO Laboratory and Advanced LIGO are funded by the United States National Science Foundation (NSF) as well as the Science and Technology Facilities Council (STFC) of the United Kingdom, the Max-Planck-Society (MPS), and the State of Niedersachsen/Germany for support of the construction of Advanced LIGO and construction and operation of the GEO600 detector. Additional support for Advanced LIGO was provided by the Australian Research Council. Virgo is funded, through the European Gravitational Observatory (EGO), by the French Centre National de Recherche Scientifique (CNRS), the Italian Istituto Nazionale di Fisica Nucleare (INFN) and the Dutch Nikhef, with contributions by institutions from Belgium, Germany, Greece, Hungary, Ireland, Japan, Monaco, Poland, Portugal, Spain.
%%%%%%%%%%%%%%%%%%%%%%%%%%%%%%%%%%%%%%%%%%%%%%%%%%%%%%%%%
%%%%%%%%%%%%%%%%%%%%%%%%%%%%%%%%%%%%%%%%%%%%%%%%%%%%%%%%%
\appendix

%%%%%%%%%%%%%%%%%%%%%%%%%%%%%%%%%%%%%%%%%%%%%%%%%%%%%%%%%
%%%%%%%%%%%%%%%%%%%%%%%%%%%%%%%%%%%%%%%%%%%%%%%%%%%%%%%%%
\section{Vetoes and Estimation of $\pastro$}
\label{Appendix:pastro}

\begin{figure}
    \centering
    \includegraphics[width=\linewidth]{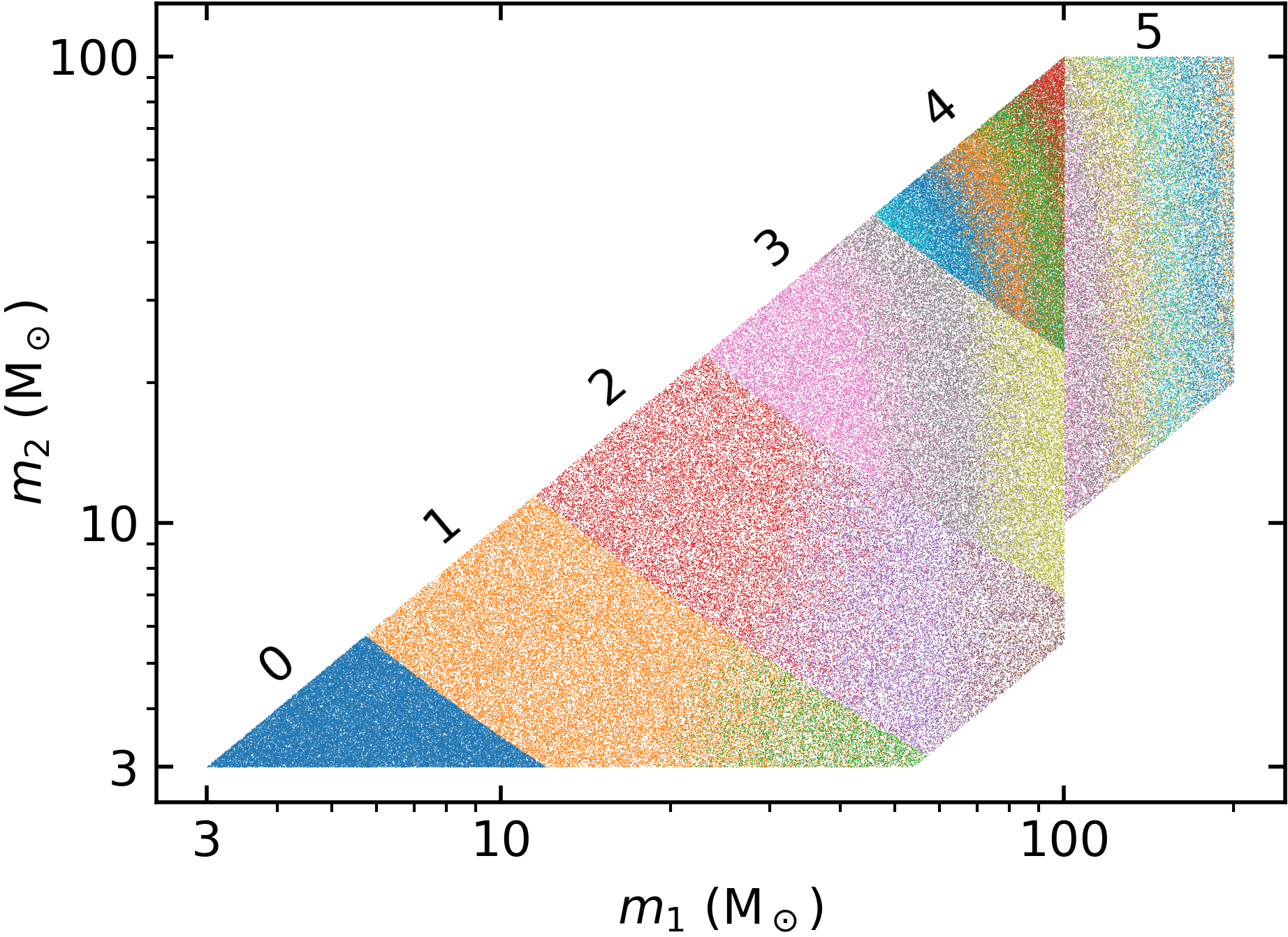}
    \caption{We use the same template bank as in our O3a analysis \cite{ias_o3a_pipeline2022} (see Ref.~\cite{ias_template_bank_PSD_roulet2019} for details on our bank construction using geometric placement). The labeled numbers denote the different banks (the first four banks are divided based on the chirp mass while the fifth bank is divided using the mass of the primary). The colors denote the different subbanks wherein the amplitude profiles remain the same while the phases of the templates could vary.} 
    %\jay{Unless some discussion is centered on this figure, I would suggest moving it to the appendix.}
    \label{fig:banks}
\end{figure}

Let us define ``trigger" to mean a template and a segment of data for which the quadrature sum of matched-filter SNR from each detector (let us call it an incoherent score) passes a certain threshold. 
A typical event will produce triggers above the follow-up threshold associated with multiple templates when the data stream is match-filtered against the template bank (shown in Fig.~\ref{fig:banks}). 
Within a segment, we pick the trigger with the highest incoherent score as the representative one for further follow-up. At the next stage, the triggers are then assigned a coherent score (see Appendix D of \cite{ias_o3a_pipeline2022}) which incorporates information about the consistency of the phases, amplitudes and arrival times between the detectors. These (coherent) scores can be used to rank triggers. However, the different templates within a bank do not represent the same amount of phase-space volume (or prior probability, see figure~18 of \cite{ias_o3a_pipeline2022}). Triggers associated with templates from low-probability regions of the bank should be down-weighted relative to those from regions with larger probability in order to improve the search's overall sensitivity to the population \cite{optimal_template_prior_popdist_dent2014}. We achieve this by using an additional template prior (see section C of \cite{ias_o3a_pipeline2022}). Finally, we combine these two pieces of information to compare the triggers within a few milliseconds of each other across sub-banks or banks. The one which has the highest value of $\Tilde{C}\equiv \log (\rm{coherent\,\, score}) + \log(\rm{template\,\, prior\,\, density})$ is selected and the event is assigned to that trigger. Different events are then ranked using a statistic which adds the rank function to $\Tilde{C}$ to account for the non-Gaussian tails from glitches (see subsection III.J of \cite{ias_pipeline_o1_catalog_new_search_prd2019}). More specifically, the final ranking score ($\Sigma$) used to rank the events is
\begin{equation}
    \Sigma = \Tilde{C} - \Big[\rho^2_H +\rho^2_L + 2 \log P(\rho^2_H|N) + 2 \log P(\rho^2_L|N) \Big]
\end{equation}
% \begin{equation}
%     \Sigma = \Tilde{C} + (\rm{Rank\,\, function} - \rm{Incoherent\,\, score})
% \end{equation}
where the quantity inside the square parenthesis becomes zero for the Gaussian noise distributions. See section~4 of \cite{Wad23_Pipeline} for further details on the justification of the above equation.

Previously, as in \citet{ias_o3a_pipeline2022}, we applied vetoes on individual triggers {\em before} maximizing the statistic $\hat{C}$ over different template banks when picking which bank to assign each trigger to. We found that this procedure led to a slight overproduction of background triggers. Here we give an argument for why that occurs. Say a candidate produced triggers in three subbanks (or banks), and their scores are $\Tilde{C}_1$, $\Tilde{C}_2$ and $\Tilde{C}_3$. Suppose further that the trigger associated with $\Tilde{C}_3$ was vetoed out when we applied our signal-consistency tests. If $\Tilde{C}_3$ were the highest among the three scores, i.e., the ``best-fit" template was not a good match to the data, we would not want to keep this candidate in our catalog. However, the maximization process would then pick one of the remaining two lower scores $\Tilde{C}_1$ and $\Tilde{C}_2$, and the candidate would enter the catalog. In this work, we switch the order: we first maximize over the scores $\Tilde{C}$ and then apply vetoes. This way all the three triggers are removed if $\Tilde{C}_3$ were the highest and got vetoed. This procedure slightly improves our background. However, we also pay the price of losing a few foreground triggers.    

The astrophysical probability ($\pastro$) of an event with ranking score $\Sigma_0$ is defined as follows;
\begin{equation}
    \pastro(\Sigma_0)= \dfrac{\dfrac{dN}{d\Sigma} (\Sigma_0 \mid H_S)}{\dfrac{dN}{d\Sigma} (\Sigma_0 \mid H_S) + \dfrac{dN}{d\Sigma} (\Sigma_0 \mid H_N)}
\end{equation}
where $H_S$ denotes the signal hypothesis and $H_N$ the noise hypothesis. Then,  $dN/{d\Sigma} (\Sigma_0|H_S)$ and $dN/{d\Sigma} (\Sigma_0|H_N)$ denote the foreground and background trigger densities at the score $\Sigma_0$, respectively. We compute $\pastro$ following the same procedure outlined in Appendix B of \citet{ias_o3a_pipeline2022}. We measure the foreground and background trigger distributions directly from the data. {We show these distributions in Fig.~\ref{fig:ranking_score_plot}.} To estimate the background, we generate an additional 2000 O3b runtimes worth of noise realizations from the O3b data using timeslides, i.e., unphysical time shifts between the data at the Hanford and Livingston detectors.

\begin{figure}[t]
    \centering
    \includegraphics[width=\linewidth]{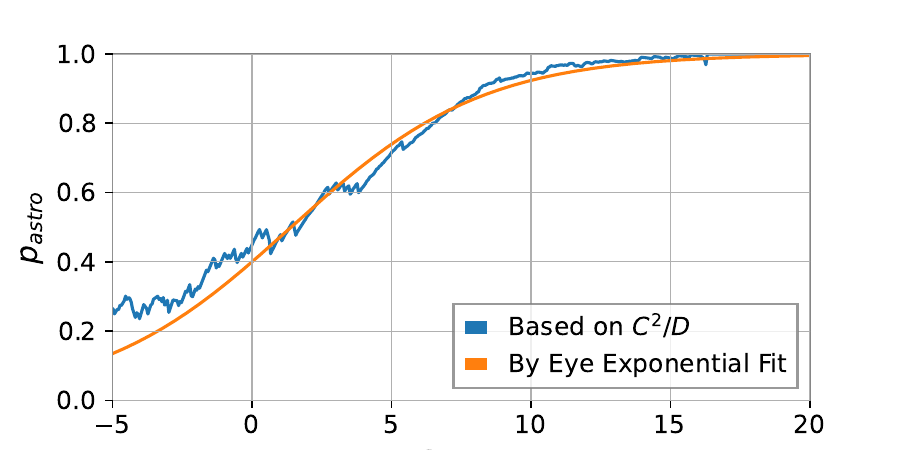}
    \caption{$\pastro$ as a function of ranking score $\Sigma$ computed based on the simple analytical model proposed in Eq. B4 of \citet{ias_o3a_pipeline2022}. The smooth line represents the fit given in Eq.~\ref{eq:pastro_fit}.}
    \label{fig:pastro_plot}
\end{figure}

In Fig.~\ref{fig:pastro_plot} we show $\pastro$ as a function of the ranking score $\Sigma$. Using the same fitting ansatz as \citet{ias_o3a_pipeline2022}, we find that
\begin{equation}
    \pastro(\Sigma) \approx \dfrac{1}{1+a e^{-\gamma \Sigma}}, \,\,\,\,\, (a, \gamma)=(1.5, 0.32)
    \label{eq:pastro_fit}
\end{equation}
provides a good fit over the relevant range. Notice that the fit coefficients are different from the ones in \citet{ias_o3a_pipeline2022}, as we could have expected.

\begin{figure*}[t]
    \centering
    \includegraphics[width=\linewidth]{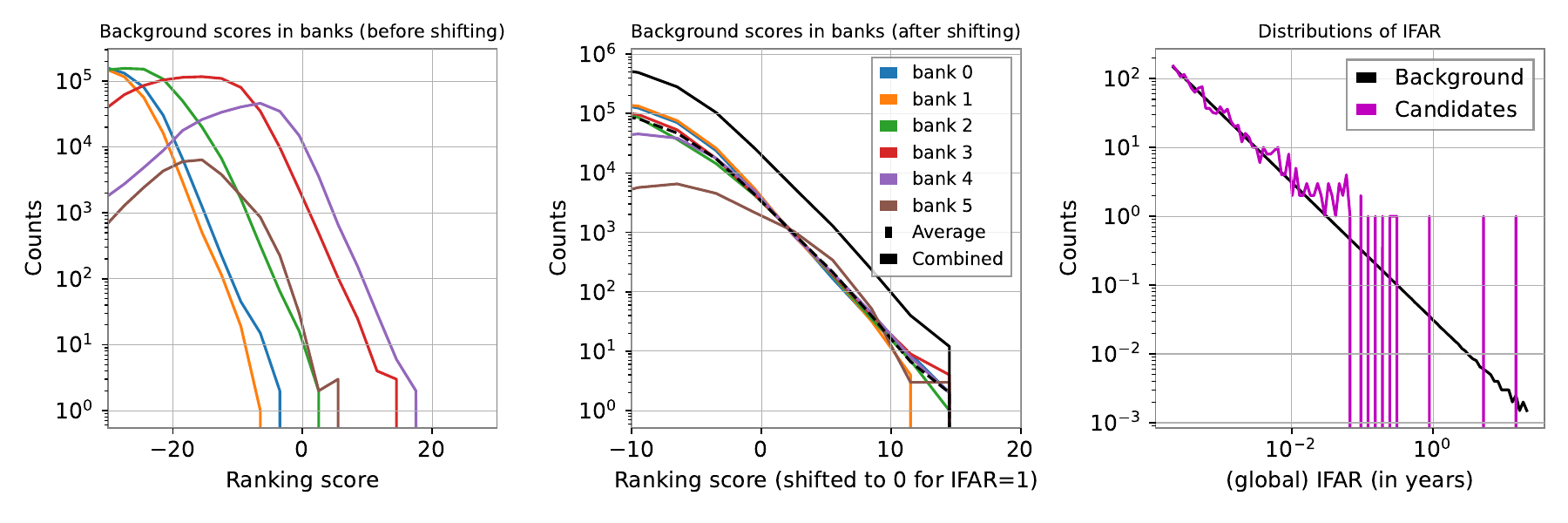}
    \caption{{\textit{Left panel}: Distribution of the ranking statistic score for the background from each bank, obtained from 2000 O3b time slides. The relative shifts between banks are arbitrary as the astrophysical prior is normalized separately within each bank. \textit{Middle panel}:
    In our search, we use a prior that equal number of astrophysical signals are expected in each of our banks. We therefore shift the scores from each bank so that they match at IFAR=1.
    % a score of zero corresponds to the expectation of one trigger during O3b for that bank. 
    Also plotted are the combined (solid black) and averaged (dashed black) distributions over the banks. We generate the combined distribution by concatenating the shifted scores of each bank.  \textit{Right panel}: The distributions of the IFAR for the background and the foreground (candidates) during the O3b run.}}
    \label{fig:ranking_score_plot}
\end{figure*}

\section{Posteriors for the new events}
\label{Appendix:posteriors}

Here we present corner plots of the posteriors for a few selected parameters for our new events: the source-frame constituent masses $m_{1}^{\rm{src}}$ and $m_{2}^{\rm{src}}$, the effective spin $\chi_{\rm{eff}}$, and the redshift $z$. The sampling prior used here is the same as in Ref.~\cite{ias_o3a_pipeline2022}, i.e., uniform in detector-frame constituent masses (as in GWTC-3 and 4-OGC), effective spin (for more details on the flat effective spin prior, see \cite{flat_chieff_prior_o1event1216_formation_channels2019} or \cite{ias_gw190521_prd2021olsen}), and comoving $VT$ (using a $\Lambda$CDM cosmology with Planck15 results \cite{cosmology_planck2015}). These samples are publicly available at \url{https://github.com/AjitMehta/new_BBH_mergers_O3b_IAS_pipeline}. The posteriors were generated using the same package $\tt{cogwheel}$ as in \cite{ias_o3a_pipeline2022} but with the most recent commit \url{https://github.com/jroulet/cogwheel/tree/c15d40ab29faf595e6dcd7951b15ccc92cdf1296} at the time when the analysis was conducted. The other details are also the same, e.g., we use 8192 live points with a log-evidence tolerance value of 0.1 for the underlying $\tt{PyMultiNest}$ sampler \cite{pymultinest}.

%The likelihood is generated with IMRPhenomXPHM waveform using the relative binning method described in \citet{relative_binning}.

\begin{figure}[H]
    \centering
    \includegraphics[width=\linewidth]{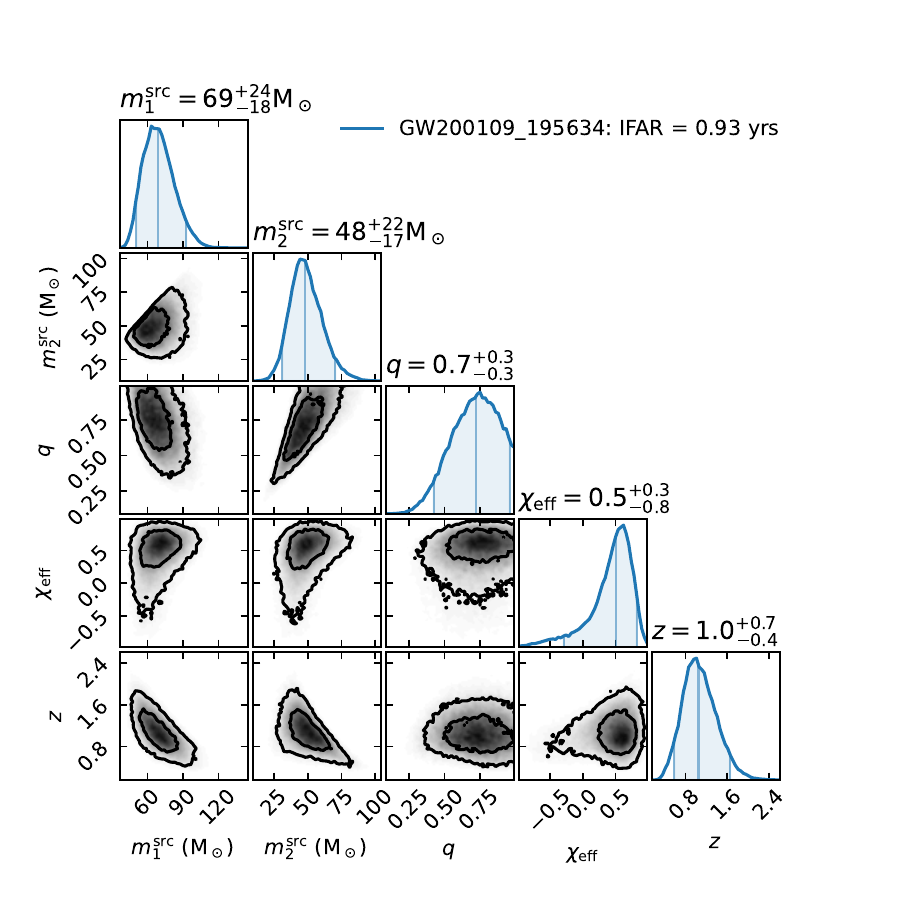}
    \vspace{-0.82cm}
    \caption{Corner plot for the event GW$200109\_195634$ with $\pastro=0.81$. The primary BH lies in UMG. This event has a noticeable contribution from the Virgo detector as well. Our future searches where Virgo data is included will thus improve its significance further. The redshift observed is highest among our newly detected events.}
    \label{fig:GW200109_195634_corner_plot}
\end{figure}

\begin{figure}[H]
    \centering
    \includegraphics[width=\linewidth]{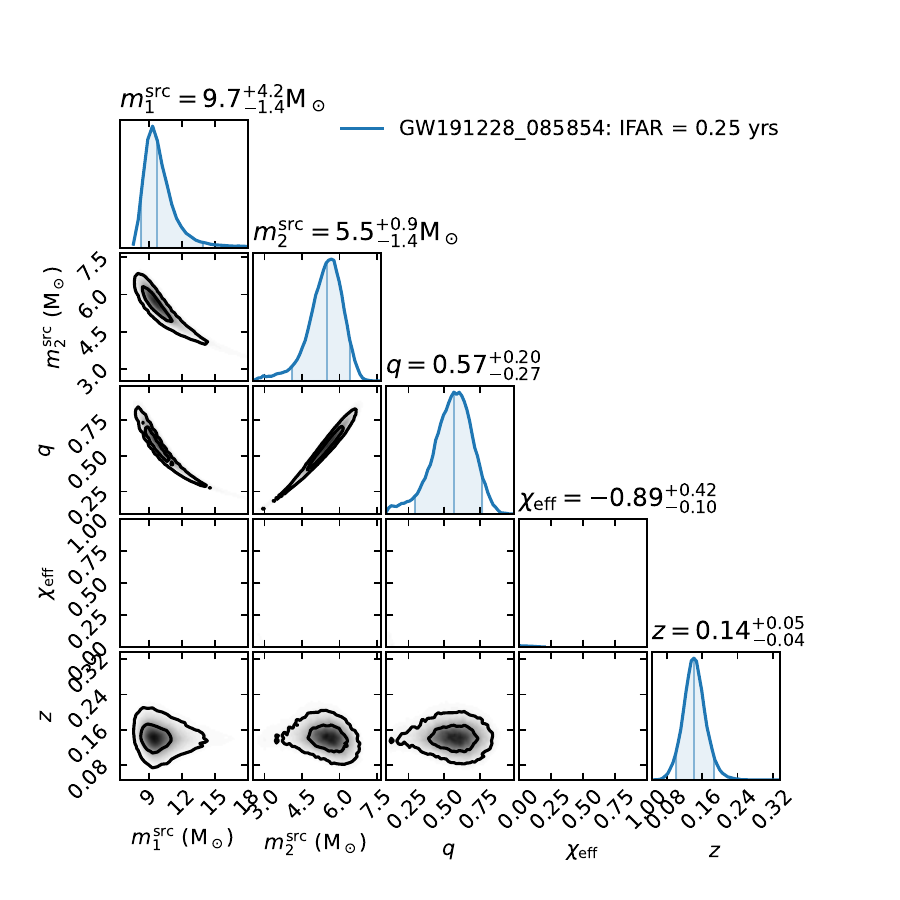}
    \vspace{-0.82cm}
    \caption{Corner plot for the event GW191228\_085854 with $\pastro=0.67$. The effective spin is highly negative. See appendix \ref{Appendix:bimodality} for discussion.}
    \label{fig:GW200109_195634_corner_plot}
\end{figure}

\begin{figure}[H]
    \centering
    \includegraphics[width=\linewidth]{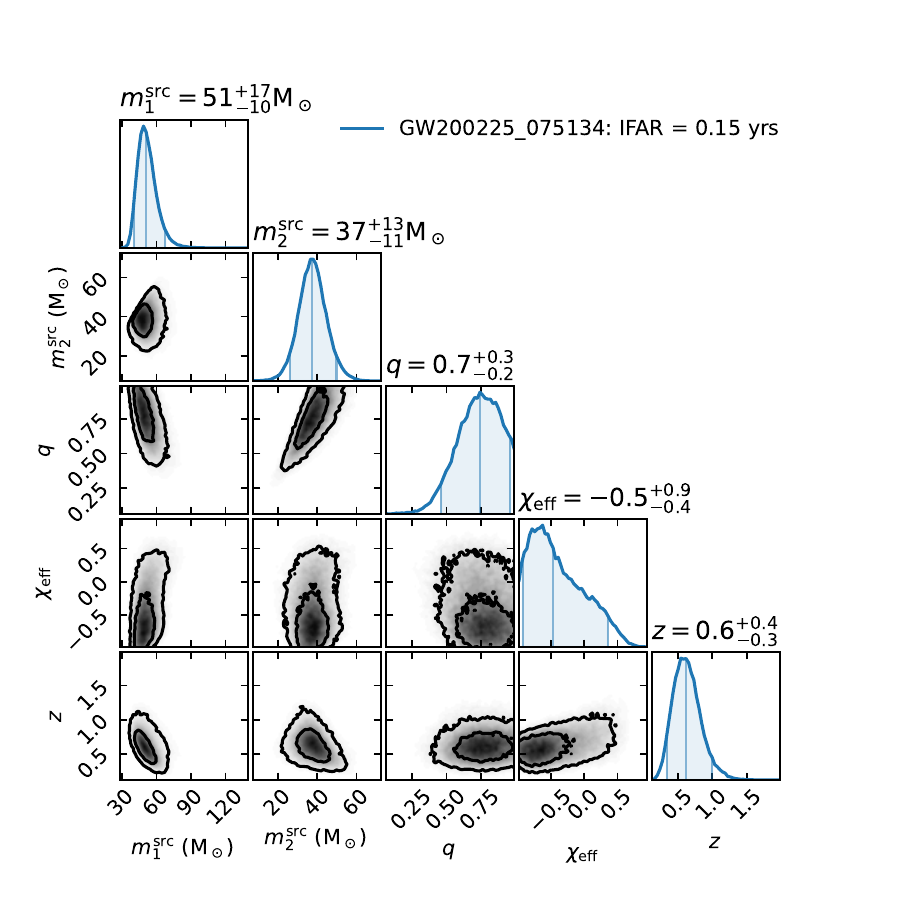}
    \vspace{-0.82cm}
    \caption{Corner plot for the event GW$200225\_075134$ with $\pastro=0.60$. This event seems to be consistent with a typical stellar BBH.}
    \label{fig:GW200225_075134_corner_plot}
\end{figure}

\begin{figure}[H]
    \centering
    \includegraphics[width=\linewidth]{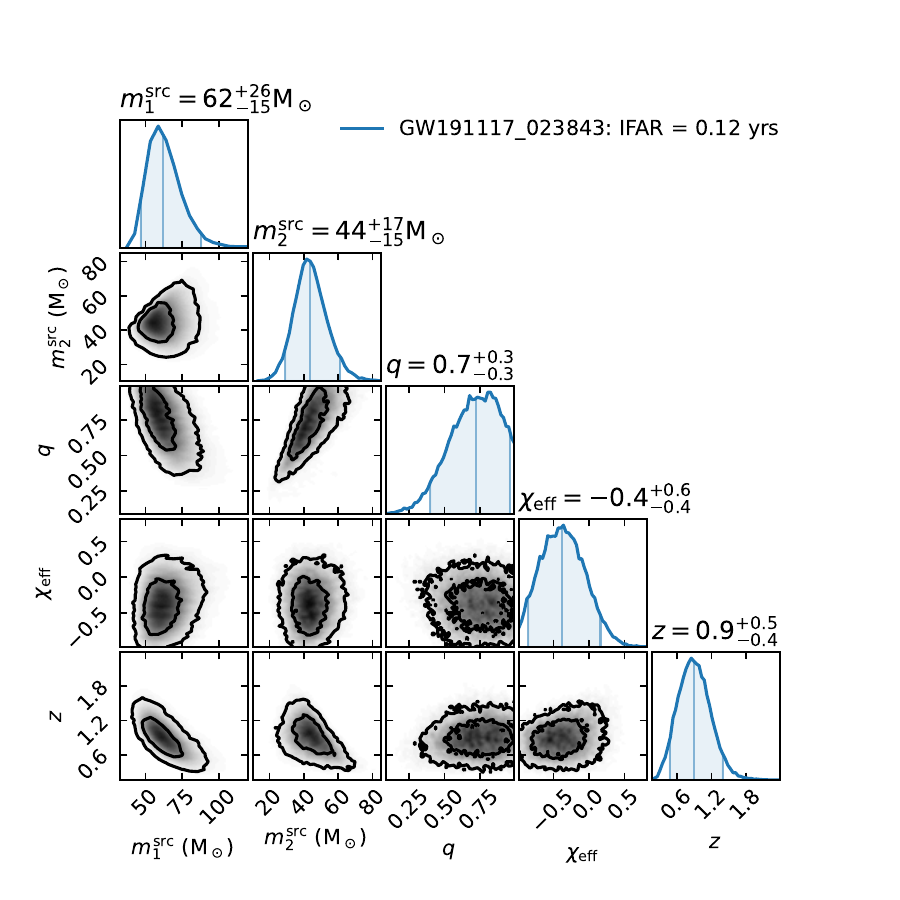}
    \vspace{-0.82cm}
    \caption{Corner plot for the event GW$191117\_023843$ with $\pastro=0.56$. The primary BH lies confidently in UMG.}
    \label{fig:GW191117_023843_corner_plot}
\end{figure}

\begin{figure}[H]
    \centering
    \includegraphics[width=\linewidth]{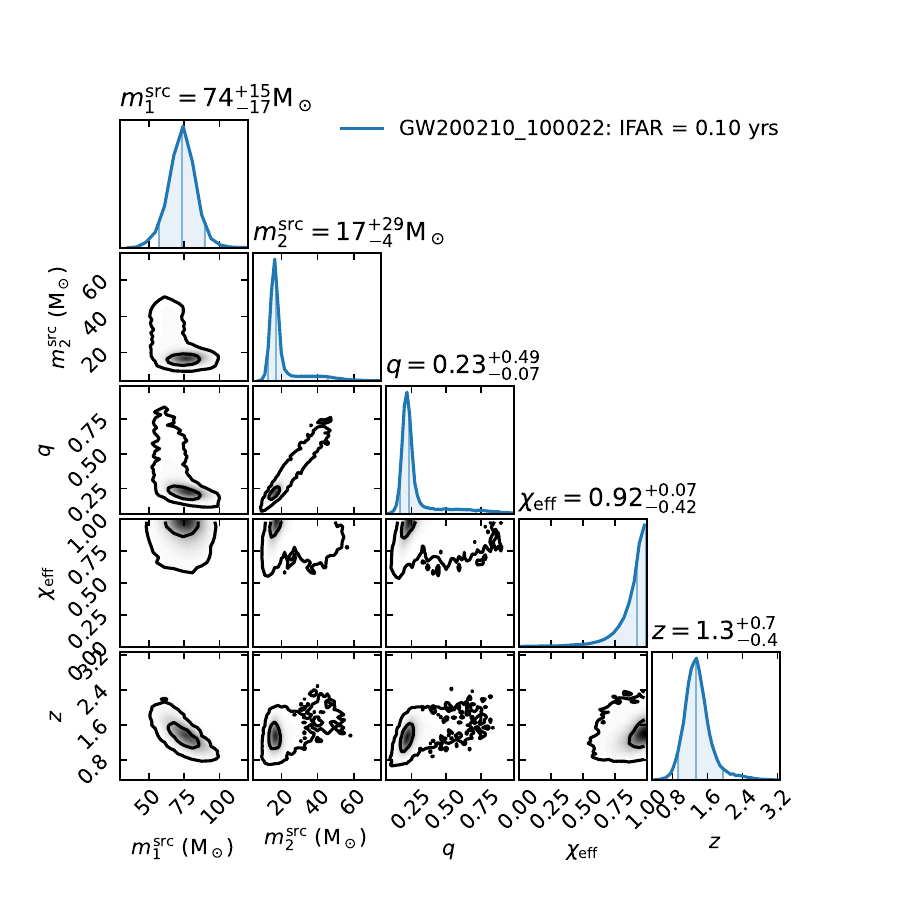}
    \vspace{-0.82cm}
    \caption{Corner plot for the event GW$200210\_100022$. It is one of the least significant events in our list of new events, with $\pastro=0.52$. The primary BH lies confidently in UMG. The effective spin is quite high, $\chi_{\rm{eff}}\gtrsim 0.5$ at $90\%$ confidence.}
    \label{fig:GW200210_100022_corner_plot}
\end{figure}

\begin{figure}[H]
    \centering
    \includegraphics[width=\linewidth]{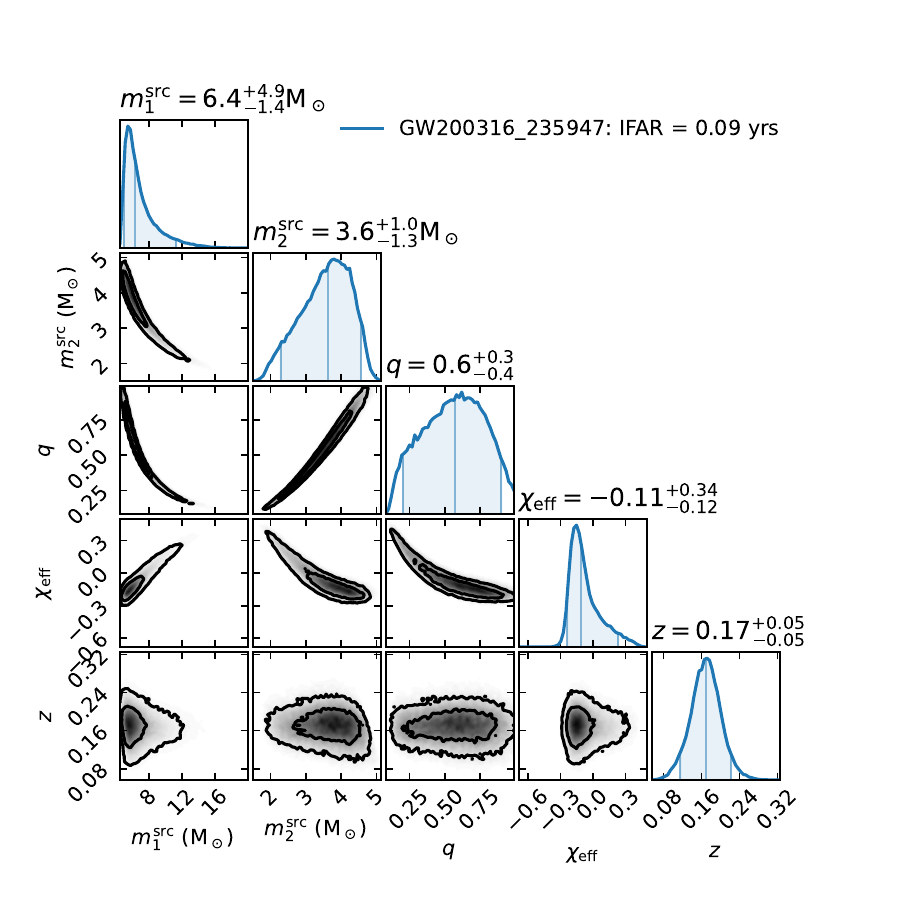}
    \vspace{-0.82cm}
    \caption{Corner plot for the most marginal event GW$200316\_235947$ with $\pastro=0.52$. The secondary constituent confidently lies in LMG. It can also be a heavy NS. The likelihood comparison between the regions $m_{2}^{\rm{src}}\lesssim 2.6\,\msun$ (NSBH) and $m_{2}^{\rm{src}}\gtrsim 2.6\,\msun$ (BBH) shows that both the NSBH and BBH solutions are equally favorable. We show the sky localization for this event in Fig.~\ref{fig:GW200316_235947_sky_loc}.}
    \label{fig:GW200316_235947_corner_plot}
\end{figure}

\section{Bi-modality in posteriors: GW191228\_085854 and the LVK analog GW200210\_092254}
\label{Appendix:bimodality}
One of the new events reported in this work, GW191228\_085854, has a posterior distribution that is bi-modal in several intrinsic parameters when we analyse it with our default spin prior that is uniform in the effective spin (See Fig. \ref{fig:GW191228_085854_corner_plot} below, in which this prior is labelled as the ``IAS prior"). Posteriors obtained using the LVK collaboration's default spin prior (isotropic in individual spins), on the other hand, do not show any such feature and pick out only the secondary mode of the result obtained using the IAS prior. The first posterior mode suggests that this system is a BBH merger with large and anti-aligned spins, while the second mode points to a BBH system with significantly asymmetric masses, but zero effective spin. 

\begin{figure}[htb]
    \centering
    \includegraphics[width=\linewidth]{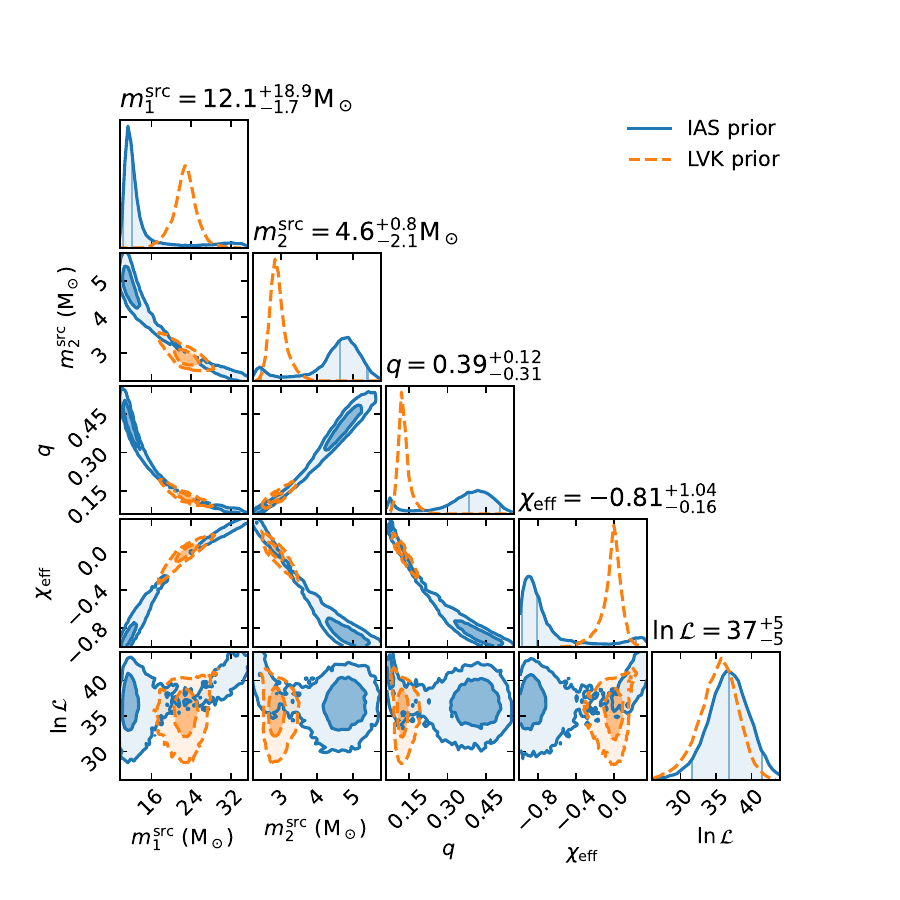}
    \vspace{-0.82cm}
    \caption{Corner plot for the LVK event GW200210\_092254. This event has very similar features in the posteriors as GW191228\_085854.}
    \label{fig:GW200210_092254_corner_plot}
\end{figure}

The fact that the posterior, and consequently the astrophysical interpretation, can differ depending on the prior choices is not new (especially for low SNR events) but this event calls for further investigation. 
We looked for analogs in the events within the LVK catalog, and found similar (i.e., low total mass) events within. 
For example, see Fig.~\ref{fig:GW200210_092254_corner_plot} for the event GW200210\_092254. It can be seen to have very similar features in the posteriors. Other low mass LVK events GW191113\_071753 and GW190917\_114630 also exhibit bi-modality in some of their intrinsic parameters. We leave detailed investigations of such systems to future work.

\begin{figure*}[htb]
    \centering
    \includegraphics[scale=0.57]{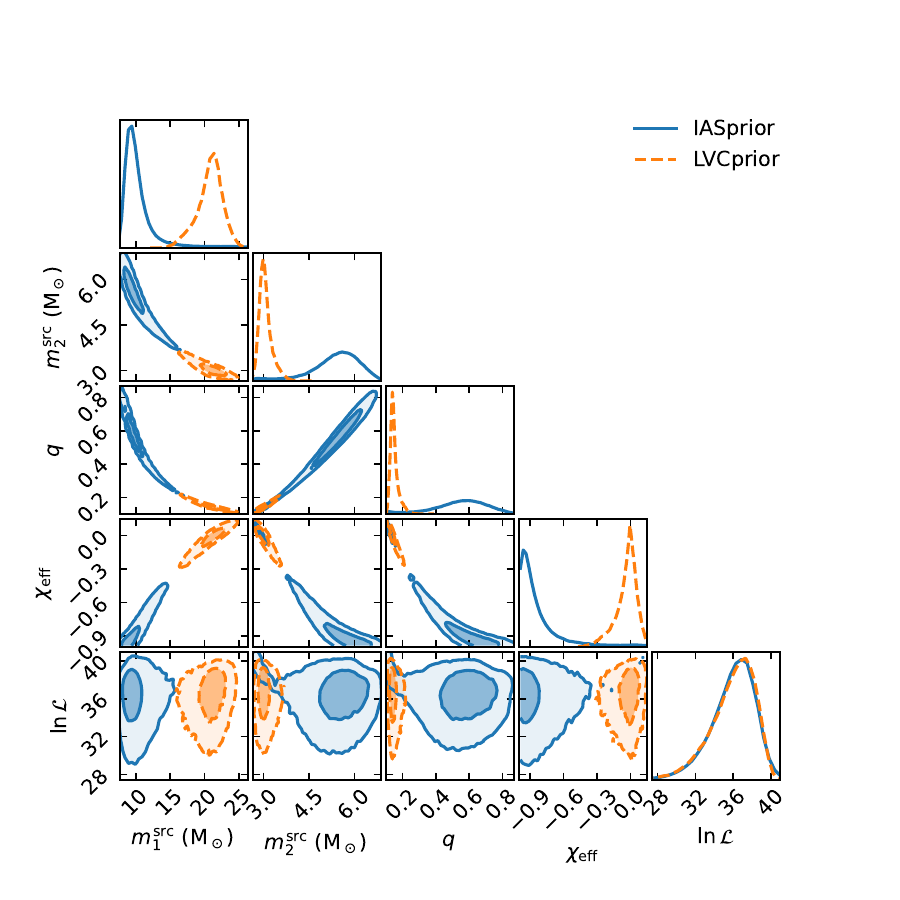}
    \includegraphics[scale=0.57]{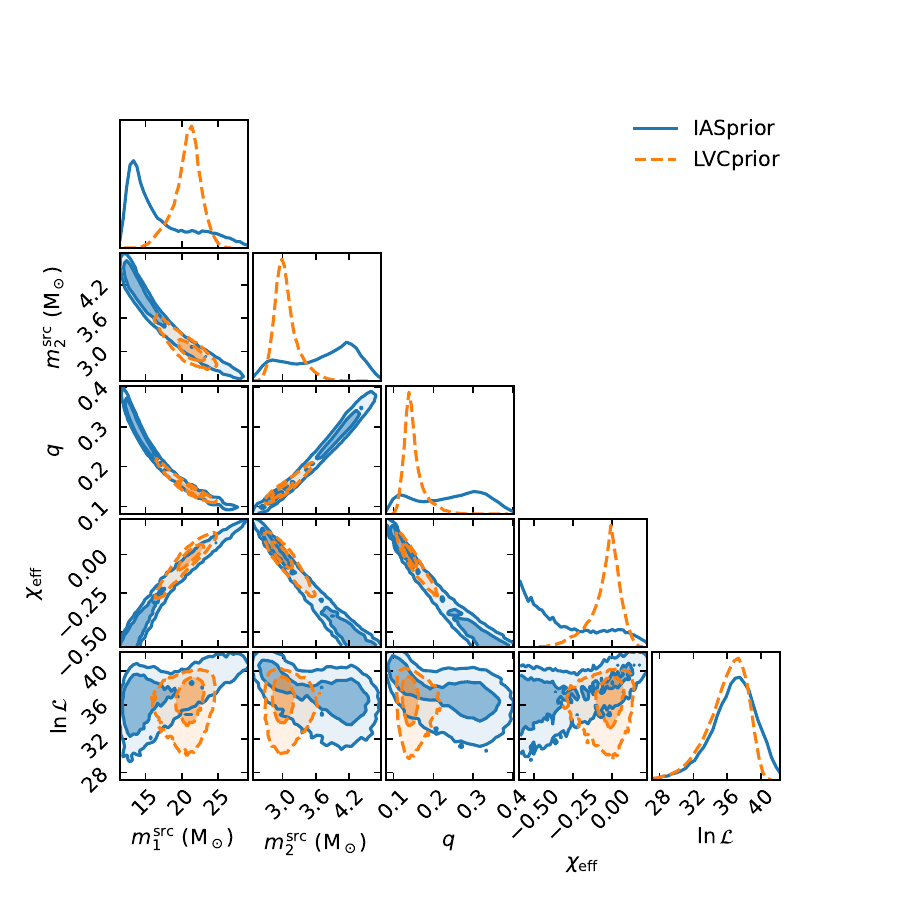}
    \vspace{-0.32cm}
    \caption{A closer look at the new event GW$191228\_085854$. \textit{Left panel:} The results obtained with a prior uniform in effective spin (IAS prior) and isotropic in the spins (LVK prior). \textit{Right panel:} Same as the left panel but with a cut in the posterior samples for the IAS prior, $\chi_{\rm{eff}}>-0.6$ (in other words, the zoomed-in version of the left panel plot near less negative $\chi_{\rm{eff}}$). The bi-modal feature in the IAS prior result is more clearly visible in the right panel plot. The LVK prior selectively picks out the second mode.}
    \label{fig:GW191228_085854_corner_plot}
\end{figure*}

%%%%%%%%%%%%%%%%%%%%%%%%%%%%%%%%%%%%%%%%%%%%%%%%%%%%%%%%%
%%%%%%%%%%%%%%%%%%%%%%%%%%%%%%%%%%%%%%%%%%%%%%%%%%%%%%%%%
%%%%%%%%%%%%%%%%%%%%%%%%%%%%%%%%%%%%%%%%%%%%%%%%%%%%%%%%%
%%%%%%%%%%%%%%%%%%%%%%%%%%%%%%%%%%%%%%%%%%%%%%%%%%%%%%%%%

\bibliographystyle{apsrev4-1-etal}
\bibliography{main}
%------------------------------------------------------------------------------

\end{document}